\documentclass[10pt,conference]{IEEEtran}
\usepackage[T1]{fontenc}
\IEEEoverridecommandlockouts
\usepackage{cite}
\usepackage{amsmath,amssymb,amsfonts}
\usepackage{algorithmic}
\usepackage{graphicx}
\usepackage{textcomp}
\usepackage{xcolor}

\usepackage{tabularx}
\usepackage{booktabs}
\usepackage{array}
\usepackage{listings}
\usepackage{xurl}
\usepackage{minted}
\usepackage{enumitem}
\usepackage{subcaption}

\def\BibTeX{{\rm B\kern-.05em{\sc i\kern-.025em b}\kern-.08em
    T\kern-.1667em\lower.7ex\hbox{E}\kern-.125emX}}
\begin{document}

\title{Don't Make Models Guess Security and Safety: Symbolic Guardrails for Domain-Specific AI Agents}

\author{}
\author{\IEEEauthorblockN{Yining Hong*, Yining She, Eunsuk Kang, Christopher S. Timperley and Christian K\"astner}
\IEEEauthorblockA{
% School of Computer Science\\
Carnegie Mellon University\\
% Pittsburgh, PA, USA\\
*yhong3@andrew.cmu.edu}}

\maketitle

\begin{abstract}
There is increasing interest in integrating AI agents that invoke tools into domain-specific commercial software, where unintended tool calls can cause serious security and safety incidents. This has drawn growing research attention, and many agent security and safety benchmarks have emerged. They implicitly shape how the community approaches security and safety. Yet existing work exhibits a blind spot: it emphasizes training-based methods and neural guardrails, which reduce the likelihood of insecure or unsafe actions but cannot guarantee their prevention. It generally overlooks opportunities for deductive, symbolic guardrails grounded in standard software engineering practices, which can provide guarantees for some security and safety requirements. Our study has three parts: (1) a systematic review of 80 agent security and safety benchmarks finding that that 85\% of benchmarks do not state verifiable requirements (61\% provide none, and 24\% give only high-level goals); (2) an applicability analysis of which security and safety requirements symbolic guardrails can and cannot enforce on $\tau^2$-Bench, CAR-bench, and MedAgentBench, finding that 74\% of requirements are symbolically enforceable and 95\% of these need only simple, low-cost checks; and (3) an empirical evaluation of symbolic guardrails on the same three benchmarks, finding that symbolic guardrails improve security and safety without sacrificing utility, and often improve it. Our work draws attention to the potential for symbolic guardrails for AI agents, suggesting them as an overlooked but practical path toward deploying domain-specific AI agents in risk-averse commercial software. We release all codes and artifacts at \url{https://github.com/hyn0027/agent-symbolic-guardrails}.
\end{abstract}

\begin{IEEEkeywords}
AI Agents, Software Security, Agent Safety, Agent Security, Symbolic Guardrails
\end{IEEEkeywords}

% \begin{figure}[ht]
%  \centerline{
%   \includegraphics[width=0.8\linewidth]{figures/AI_agents.pdf}}
%  \caption{Overview of the AI agent workflow. The LLM interacts with the user, performs reasoning, and invokes tools.}
%   \label{fig:overview_of_agent}
% \end{figure}

\begin{figure}[t]
\centering
  \includegraphics[width=0.94\linewidth]{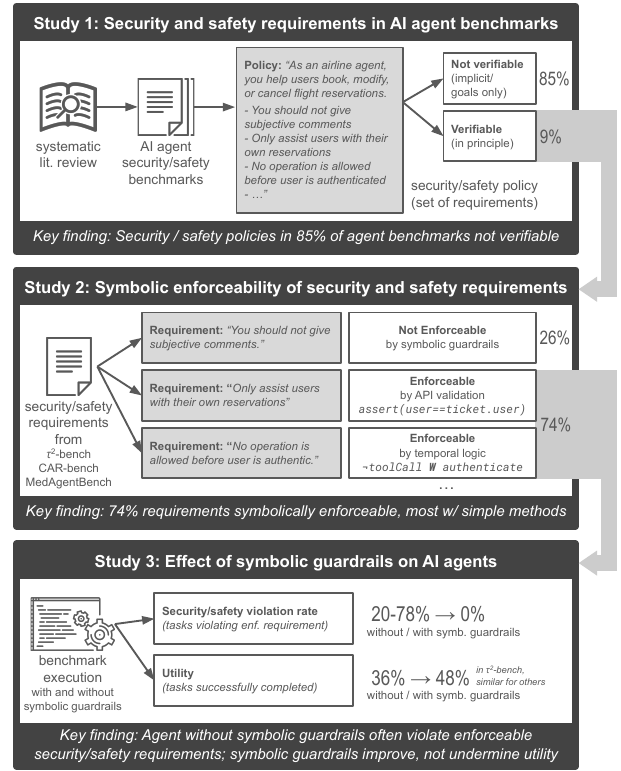}
 \caption{Overview of the three-part study.}\label{fig:study_overview}
\end{figure}

\section{Introduction}\label{sec:intro}

AI agent security and safety have become an active research area~\cite{ma2026safety,10.1145/3716628}. This paper argues that current agent security and safety benchmarks exhibit a large blind spot: they emphasize alignment (training the LLM to internalize human preferences) and neural guardrails (auxiliary LLMs that screen agent traces), so that agents meet vague \textit{common-sense} security and safety expectations. However, these benchmarks overlook effective and widely applicable symbolic guardrails grounded in standard software engineering practices, such as precondition checks on tool calls, schema enforcement, and least-privilege design. Benchmarks do not merely measure progress; they direct it~\cite{dehghani2021benchmarklottery, raji2021ai}. By enumerating what counts as success and failure, benchmarks implicitly define the problem the community optimizes for. The security and safety requirements that benchmarks encode thus serve as the operational definition of a \textit{secure} or \textit{safe} agent, and the defenses that the community builds are the ones these benchmarks reward. 

However, current benchmarks do not capture the needs of risk-averse commercial applications.
AI agents are increasingly deployed not just as personal assistants but to automate business functions at scale, such as customer support~\cite{sierra2026}, healthcare~\cite{hippocratic2026}, and software development~\cite{cursor}, yet many organizations are wary of risks. Individual users may accept some personal risk when adopting assistants such as OpenClaw~\cite{openclaw2026}, but when agents act on behalf of a company, a single incident can have serious consequences, such as compromising customer accounts~\cite{insta_agent_incident} or authorizing transactions that violate company policies~\cite{Cecco2024AirCanadaChatbot}; such risks fall centrally on the system operator. Agents are more secure and safer when stronger LLMs and neural guardrails are used: in our study, upgrading the LLM from GPT-4o to GPT-5 cuts the rate of runs violating security and safety requirements in $\tau^2$-Bench~\cite{barres2025tau} from 52\% to 20\%. Yet it is unclear how continued progress of this kind can instill enough confidence to deploy a commercial system when a single incident, such as a leaked patient database or an irreversible transaction, can cause severe damage~\cite{1335256}. The relevant question in such risk-averse settings is therefore not how an agent \textit{usually} behaves, but whether certain actions can be \textit{ruled out}, even under an adversary; this is a guarantee that neural defenses, inductive by definition, cannot provide.

In this work, we examine whether current benchmarks, as a proxy for the community's attention, support building agents for high-assurance settings. As in \figurename~\ref{fig:study_overview}, we conducted a \textbf{three-part study}: a systematic literature review of 80 agent security and safety benchmarks (Sec.~\ref{sec:part1_collecting_policies}), an applicability analysis of which requirements are (not) symbolically enforceable across three benchmarks (Sec.~\ref{sec:part2_analyzing_policies}), and an empirical evaluation of how symbolic guardrails affect agent security, safety, and utility (Sec.~\ref{sec:part3_experiments}). We reached three major findings: 

\textbf{First, 85\% of security and safety benchmarks do not state verifiable requirements.} Of the 80 identified benchmarks, 61\% provide no security or safety requirements on agent behaviors, expecting agents to follow unstated \textit{common-sense} norms; another 24\% provide only high-level goals with operative terms undefined, such as ``If user requests anything unsafe, do not fulfill''~\cite{zhou2026safeproevaluatingsafetyprofessionallevel} with \textit{unsafe} unexplained. Engineers cannot assure what they cannot specify, so most benchmarks offer limited insight for high-assurance settings. Notably, the few benchmarks with verifiable requirements all target \textit{domain-specific} agents, where bounded tasks and tools make requirements enumerable, suggesting that verifiable policies are feasible precisely in the commercial applications we target.

\textbf{Second, among benchmarks that do state verifiable security and safety requirements, most are symbolically enforceable using classical software engineering mechanisms that the agent literature has largely overlooked.} Confidentiality and integrity, for example, have long been enforced through \textit{symbolic}, deductive mechanisms such as access control~\cite{sandhu2000nist}, input validation~\cite{Halfond2006ACO}, and information-flow control~\cite{10.1145/360051.360056}, which afford formal, logic-based assurances and build on principles such as least privilege and complete mediation~\cite{viega2001building}. The agent literature, by contrast, emphasizes neural guardrails such as LLM-as-a-judge~\cite{luo-etal-2025-agrail, chennabasappa2025llamafirewallopensourceguardrail, rebedea-etal-2023-nemo}, which use probabilistic, inductive methods to enforce rules, some of which could instead be enforced symbolically for a formal guarantee. For example, we observed that in $\tau^2$-Bench, an airline agent can erroneously cancel an already-flown flight~\cite{yao2024tau}, because the corresponding rule is passed through a prompt rather than enforced symbolically in the agent harness. In our study, we considered six types of symbolic guardrails. We found 74\% of the analyzed requirements are symbolically enforceable, of which 95\% need only simple, low-cost checks, predominantly API validation, rather than more complex methods such as information-flow tracking~\cite{costa2025securingaiagentsinformationflow} or temporal logic~\cite{wang2025agentspeccustomizableruntimeenforcement}. We argue that these mechanisms are critical for the path toward deploying agents in risk-averse commercial applications.

\textbf{Third, enforcing these requirements symbolically improves agent utility rather than eroding it.} A common objection to agent guardrails is that they over-constrain the agent, improving security and safety at the cost of task success~\cite{tan2025equilibraterlhfbalancinghelpfulnesssafety, dong2025safeguarding}. Our experiments show the opposite. Symbolic guardrails eliminate a large class of violations of enforceable requirements (which occur in 20\% to 78\% of tasks when the requirements are given only in the prompt and judgment is left to the model), while maintaining or improving the task completion rate. One explanation is that symbolically blocked actions give the agent clear feedback to reflect and retry with a safe alternative, rather than finishing with the wrong result.

% As in \figurename~\ref{fig:study_overview}, we support our findings with three studies: a systematic literature review of 80 agent security and safety benchmarks (Sec.~\ref{sec:part1_collecting_policies}), an applicability analysis of which requirements symbolic guardrails can and cannot enforce across three benchmarks (Sec.~\ref{sec:part2_analyzing_policies}), and an empirical evaluation of how symbolic guardrails affect agent security, safety and utility (Sec.~\ref{sec:part3_experiments}). 
Overall, we contribute \textbf{the first systematic characterization} of whether and how agent benchmarks specify verifiable security and safety requirements, a mapping from these requirements to the symbolic mechanisms that can enforce them, and evidence that such enforcement does not degrade agent utility. These results suggest that symbolic guardrails are an overlooked but practical path toward deploying domain-specific agents in risk-averse commercial applications.

\section{Background and Related Work}\label{sec:background_related_works}

\subsection{Security and Safety Requirements}\label{sec:definingsafety}

The terms \textit{security} and \textit{safety} are often used loosely across fields, by both practitioners and researchers. This looseness is central to our study: we examine what security and safety expectations are encoded in current agent benchmarks, and the imprecision of these terms is itself part of our findings.

At a high level, security and safety requirements both aim to prevent undesirable outcomes, but differ in the source of risk they address. \textit{Security requirements} constrain how a system behaves in the presence of an adversary, and conventionally concern confidentiality, integrity, and availability. \textit{Safety requirements} constrain the harm a system may cause to its users and environment, under accidental or environmental faults rather than an adversary; they require the system to stay within its intended operating envelope and fail only into benign states (e.g., not deleting files or issuing destructive commands).

In commercial settings, there are often \textit{application-specific constraints} such as ``refunds above \$50 require approval.'' Such a constraint can be violated by an adversary's attack or by an honest mistake; either may cause financial or reputational harm to the system owner or its users. Whether we call the constraint a safety or security requirement depends on the source of risk we consider, not on how the constraint is written.

We deliberately do \textit{not} adopt these definitions of security and safety as a filter on what we study. Applying them to current benchmarks yields limited insight: the expectations of agent behavior encoded in benchmarks are rarely stated explicitly, and the sources of risk they address often overlap, so the same requirement often protects against both mistakes and adversaries. Forcing each requirement into one bin obscures more than it clarifies, and this ambiguity is itself one of our empirical findings. What the requirements we study \textit{do} share is a common form: instructions the agent is meant to follow in order to prevent a specific undesirable outcome, whether that outcome is data exfiltration, a destructive action, or a violation of an organizational policy that could lead to harm. We refer to this broader notion of a harm-preventing expectation collectively as \textit{security and safety requirements}.

\subsection{AI Agents}\label{sec:related_works_agents}

\begin{figure}[t]
  \centering
  \subfloat[General-purpose agents.\label{fig:general-purpose_agent}]{%
    \includegraphics[width=0.48\linewidth]{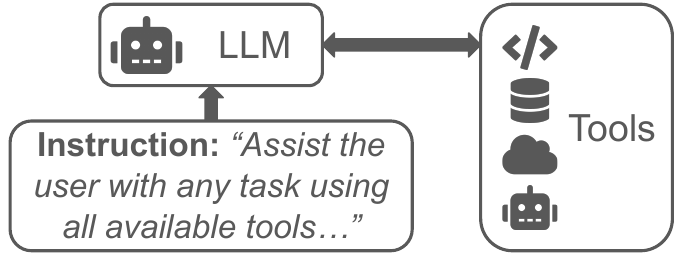}%
  }
  \hfill
  \subfloat[Domain-specific agents.\label{fig:domain_specific_agent}]{%
    \includegraphics[width=0.48\linewidth]{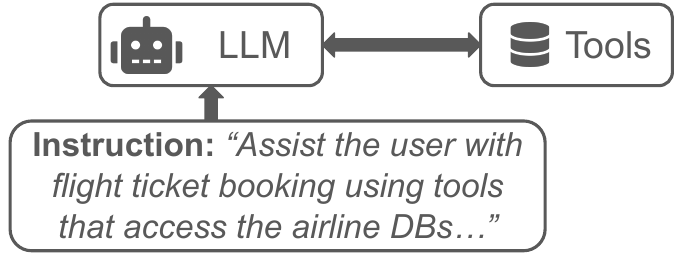}%
  }
  \caption{Two types of AI agents.}
  \label{fig:general_and_domain_specific agent}
\end{figure}

Over the past few years, large language models (LLMs) have advanced beyond single-turn text generation, gaining capabilities for context understanding, reasoning, and interaction with the environment through external tool use~\cite{schick2023toolformer}. They now serve as the core of AI agents, from the earlier retrieval-augmented generation paradigm~\cite{lewis2020retrieval} to more recent ones such as ReAct~\cite{yao2023react} and Reflexion~\cite{Reflexion}.

AI agents lack a standard definition; we focus on LLM-based agents with tool-use capabilities. The agent interacts with the user iteratively: at each step, the LLM decides the next action, either invoking tools or responding with text. Tool outputs are appended to the agent context, conditioning later actions. Tools may include APIs, command-line interfaces, or other agents. Agents autonomously decide when to invoke a tool, which to invoke, and with what arguments.
To make tools reusable across agents and to facilitate invocation, standardized formats for describing tools' purposes and schemas have been widely adopted, including Model Context Protocol (MCP)~\cite{MCP_2024} and the Agent2Agent (A2A) Protocol~\cite{A2A_2025}.

% Advances in AI agents have generated strong interest in applying them to a wide range of tasks. 
AI agents are now deployed across diverse settings.
Following prior work~\cite{lei2026offtopicevallargelanguagemodels}, we distinguish general-purpose from domain-specific agents by the scope of tasks they support. General-purpose agents (\figurename~\ref{fig:general-purpose_agent}) are designed to assist users with an open-ended set of tasks and are equipped with broadly applicable tools. Examples include ChatGPT Agent mode~\cite{Chatgpt_agent} and Microsoft Copilot integrated with Windows~\cite{copilot_windows}, which can observe the screen and control the keyboard and mouse. We also consider CLI agents such as Claude Code~\cite{anthropic_claude_code} and OpenClaw~\cite{openclaw2026} as general-purpose, since they support a wide range of tasks with broad tool-use capabilities, including file editing, command-line execution, and internet access. In contrast, domain-specific agents (\figurename~\ref{fig:domain_specific_agent}) are designed for a fixed scope of tasks, with access only to task-relevant tools. For instance, a customer support agent for airline reservations may only access reservation-related tools such as ticket lookup or refund~\cite{barres2025tau}, while a Customer Relationship Management (CRM) agent~\cite{huang2026crmarenapro} may interact only with CRM-internal workflows. Neither may access the internet or execute arbitrary code. These two categories differ in their security and safety considerations, as we discuss in Sec.~\ref{sec:part1_discussion}.

\subsection{AI Agent Guardrail: Neural versus Symbolic}\label{sec:related_works_guardrails}

Although AI agents are powerful and widely adopted~\cite{sierra2026, cursor, hippocratic2026, openclaw2026}, they may use tools to interact with environments in unintended ways, raising security and safety concerns~\cite{ma2026safety, 10.1145/3716628} (Sec.~\ref{sec:definingsafety}). To prevent agents from taking insecure or unsafe actions, researchers have explored various methods.

One line of research trains the underlying LLM supporting the AI agent to be secure and safe, embedding these properties in the model itself rather than enforcing them externally. Researchers have explored post-training alignment, particularly supervised fine-tuning and reinforcement learning from human feedback (RLHF)~\cite{ouyang2022training, bai2022hh, stiennon2020learning, glaese2022improvingalignmentdialogueagents}, and methods that replace human feedback with AI-generated signals~\cite{bai2022constitutional, lee2023rlaif, yuan2024selfrewarding}. Other work explored adversarial data collection and risk-focused active learning~\cite{perez2022red, ganguli2022reduceharms}. These methods reduce the likelihood of insecure or unsafe actions, but because LLMs are inherently probabilistic, they cannot \textit{guarantee} that agents will never violate specific security or safety requirements.

Another line of research introduces runtime guardrails built \textit{around} the LLM. We distinguish two kinds: \textit{neural} guardrails (inductive) and \textit{symbolic} guardrails (deductive).

Neural guardrails make decisions inductively, most commonly via the \textit{LLM-as-judge} paradigm: judge LLMs that are independent of the agent monitor (1) agent inputs and outputs for insecure or unsafe content, such as prompt injection and sensitive data, or (2) agent actions to assess their security and safety. For example, AGrail~\cite{luo-etal-2025-agrail} uses LLMs to generate and apply safety checks; LlamaFirewall~\cite{chennabasappa2025llamafirewallopensourceguardrail} uses ML classifiers to detect prompt injection and LLMs for misalignment; RTBAS~\cite{zhong2025rtbasdefendingllmagents} combines an LLM with a probabilistic saliency screener for information-flow analysis; and Liu et al.~\cite{299563} use LLMs to detect prompt injection.
Some work aims to provide deterministic guardrails, but still relies on LLMs to generate guardrail specifications or to execute them. These approaches, therefore, remain probabilistic and cannot provide guarantees. For example, \texttt{GuardAgent}~\cite{xiang2025guardagentsafeguardllmagents} uses LLMs to generate guardrail code; NeMo Guardrails~\cite{rebedea-etal-2023-nemo} provides programmable guardrails while execution involves judge LLMs; \textsc{ShieldAgent}~\cite{shieldagent} relies on LLMs to retrieve and execute rule-based policy checks; \textsc{AgentGuardian}~\cite{abaev2026agentguardianlearningaccesscontrol} uses LLMs to generate control policies. 
A key limitation of neural guardrails is their inductive nature: because their generation or execution depends on LLMs, their behavior cannot be formally reasoned about. They reduce the likelihood of unintended actions, but cannot guarantee an agent will never violate a security or safety requirement, which may leave deployment risk above acceptable levels in commercial settings.

%~\cite{G-Sageguard} uses GNN to detect and mediate MAS attacks.

In contrast, traditional software engineering uses symbolic, deductive enforcement techniques to provide guarantees, such as input sanitization against SQL injection~\cite{Halfond2006ACO}, information-flow control against sensitive data leakage~\cite{10.1145/360051.360056}, and access control against unauthorized access~\cite{sandhu2000nist}. A few recent studies have explored these mechanisms as symbolic guardrails for AI agent security and safety. For example,
\textsc{AgentSpec}~\cite{wang2025agentspeccustomizableruntimeenforcement}, \textsc{Agent-C}~\cite{kamath2025enforcingtemporalconstraintsllm}, and \textit{Maris}~\cite{cui2026marisformallyverifiableprivacy} use temporal logic to constrain tool-call orders; 
Progent~\cite{shi2025progentprogrammableprivilegecontrol} defines privilege control policies using domain-specific languages; Doshi et al.~\cite{doshi2026verifiablysafetooluse} explore temporal logic and information-flow control with formal models;
\texttt{PFI}~\cite{kim2025promptflowintegrityprevent} detects unsafe data flows to prevent privilege escalation; 
\textsc{Fides}~\cite{costa2025securingaiagentsinformationflow} uses information-flow control to track confidentiality and integrity;
PCAS~\cite{palumbo2026policycompilersecureagentic} likewise uses information-flow control for provenance tracking;
and $f$-secure~\cite{wu2024systemleveldefenseindirectprompt} and \texttt{CaMeL}~\cite{debenedetti2025defeatingpromptinjectionsdesign} tackle prompt injection attacks by separating control flow from data flow. 

We believe symbolic guardrails offer a practical path toward the guarantees needed to deploy AI agents in risk-averse commercial settings. Although each symbolic guardrail enforces a specific class of security or safety requirements, it is unclear whether existing symbolic guardrails cover the requirements arising in practical deployments. Moreover, there is a concern that guardrails may reduce agent utility (the ability to complete tasks) as they restrict agent behavior~\cite{tan2025equilibraterlhfbalancinghelpfulnesssafety, dong2025safeguarding}. This paper provides \textbf{the first empirical investigation} of these questions: to what extent existing symbolic guardrails apply to agent security and safety requirements, and how they affect utility.

%~\cite{agentguard}: dynamic probabilistic assurance, runtime verification given prob of whether current state will succeed/safe, but how to measure success/safety is remained tbd

% \subsection{Evaluating Security or Safety for AI Agents}
% the datasets/benchmarks

\section{Study 1: Verifiability of Security and Safety Requirements Evaluated in Agent Benchmarks}\label{sec:part1_collecting_policies} 

Benchmarks do not merely measure progress; they shape it~\cite{dehghani2021benchmarklottery,raji2021ai}. By defining success and failure, benchmarks implicitly define the problem researchers optimize for. Agents are routinely tuned and compared on security and safety benchmarks, giving them an outsized role: the requirements they encode become, in effect, the operational definition of a \textit{safe} or \textit{secure} agent. Yet in risk-averse deployments, we can only reason about requirements specified in verifiable form. This motivates our research question: \textit{which security and safety requirements do agent benchmarks evaluate, and how are these requirements specified?}

To answer this, we performed a systematic literature review of 80 agent security and safety benchmarks and examined the \textit{sets of requirements} they define, which we call \textit{policies} following prior convention~\cite{yao2024tau, barres2025tau}. We classify them along a spectrum from unstated to task-specific rules, finding that most benchmarks never state the requirements they test.

% To answer this question, we conduct a systematic literature review of 80 agent security and safety benchmarks and examine the requirements they evaluate. We classify these requirements along a spectrum from unstated to task-specific rules, finding that most benchmarks never state the requirements they test.
%Framed positively, these benchmarks evaluate an agent's common sense; framed cynically, they ask the agent to win a guess-my-mind game whose rules are revealed only by passing or failing. 

% To answer the research questions, we conduct a three-part study. First, in this section, we address RQ1 by collecting agent security and safety benchmarks and identifying the requirements they evaluate. Second, in Sec.~\ref{sec:part2_analyzing_policies}, we address RQ2 and RQ3 by analyzing which of these requirements are amenable to symbolic guardrails. Third, in Sec.~\ref{sec:part3_experiments}, we address RQ4 by evaluating the impact of symbolic guardrails on agent security, safety, and utility.

\subsection{Research Method}\label{sec:part_1_method}

\subsubsection{Identifying Benchmarks} 
\begin{figure}[t]
\centering
  \includegraphics[width=0.8\linewidth]{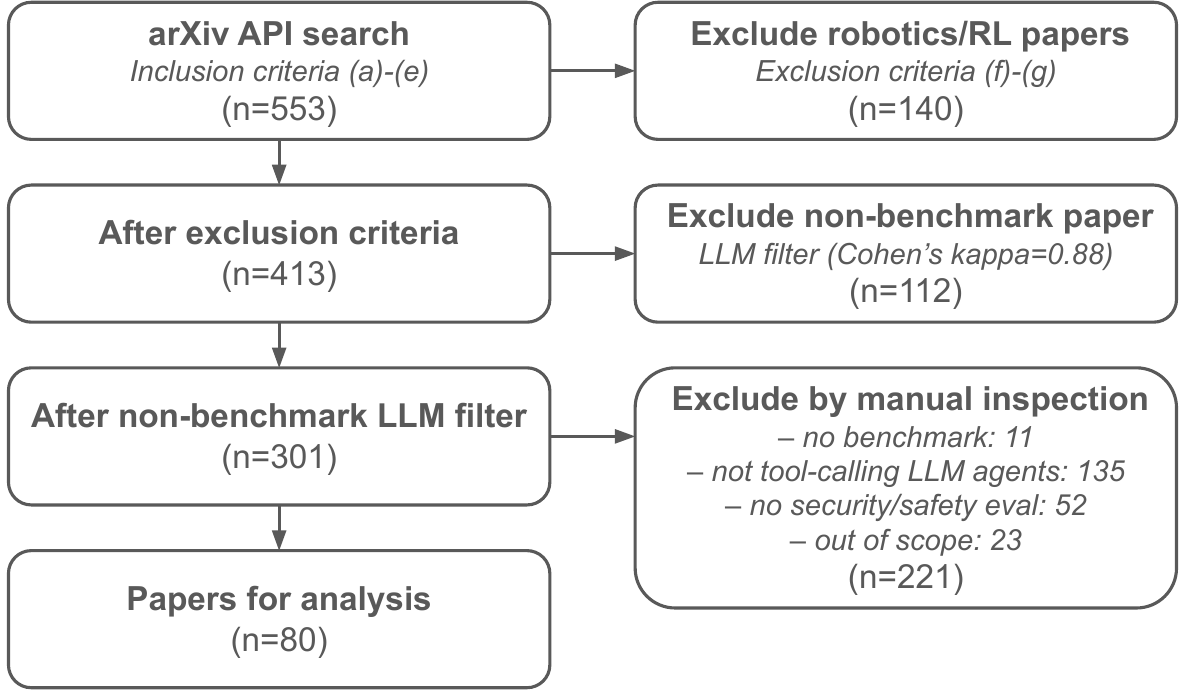}
 \caption{PRISMA-style flow of the systematic literature review.}
  \label{fig:prisma_chart}
\end{figure}

Following established guidelines~\cite{kitchenham2007guidelines}, we assembled our corpus through a systematic literature review. The process is illustrated in \figurename~\ref{fig:prisma_chart}.

\textbf{Search Criteria.}
We target papers that (1) propose a benchmark, (2) evaluate tool-calling LLM-based agents, and (3) evaluate agent security or safety. We use the arXiv API as our search interface for two reasons: AI agent papers typically appear on arXiv well before formal publication, and arXiv's subject metadata (e.g., cs.SE) enables filtering to keep the candidate set tractable. Because LLM-based agents emerged only after ChatGPT's release~\cite{chatgpt} in late 2022, we limit the search period to Jan.\ 2022 through Mar.\ 2026.

Following prior work~\cite{felizardo2016using}, we first assemble a seed set of 15 papers known to the authors. By analyzing this seed set, we derive the following inclusion criteria: (a) title or abstract contains at least one whole word from \{\textit{bench, benchmark, dataset, framework}\}; (b) title contains at least one substring from \{\textit{eval, assess, bench, dataset}\}; (c) title or abstract contains the whole word \textit{agent}; (d) title or abstract contains at least one whole word from \{\textit{safety, security, privacy, confidentiality, policy, risk, attack}\}; and (e) the paper is cross-listed in at least one of the arXiv categories cs.AI, cs.CL, or cs.LG.

% \begin{itemize}[leftmargin=*] % if we don't have enough space maybe write those in text not itemize list
%     \item Title or abstract contains at least one of the whole words: \emph{``bench''}, \emph{``benchmark''}, \emph{``dataset''}, or \emph{``framework''}.
%     \item Title contains at least one of the substrings:  \emph{``eval''}, \emph{``assess''}, \emph{``bench''}, or \emph{``dataset''}.
%     \item Title or abstract contains the whole word \emph{``agent''}.
%     \item Title or abstract contains at least one of the whole words: \emph{``safety''}, \emph{``security''}, \emph{``privacy''}, \emph{``confidentiality''}, \emph{``policy''}, \emph{``risk''}, or \emph{``attack''}.
%     \item Paper is cross-listed in at least one of the arXiv categories: cs.AI (Artificial Intelligence), cs.CL (Computation and Language), or cs.LG (Machine Learning).
% \end{itemize}

\begin{table*}[t]
  \caption{Taxonomy of AI Agent Security and Safety Policies, Classified by Verifiability and Scope}
  \label{tab:policy_types}
  \centering
  \begin{tabular}{lllp{3.9cm}p{8.5cm}}
    \toprule
    \textbf{Specificity} & \textbf{Scope} & \textbf{Verifiable} & \textbf{Explanation} & \textbf{Example} \\
    \midrule
    No Policy & - & -
      & No security/safety guidance.
      & - \\
    Goal-Setting & Agent & No
      & High-level security/safety goals governing agent behavior.
      & \textit{``If user requests anything unsafe, do not fulfill''}~\cite{zhou2026safeproevaluatingsafetyprofessionallevel}; \textit{``If query asks private information, decline to answer''}~\cite{huang2026crmarenapro}; \textit{unsafe} and \textit{private} are unexplained. \\
    Verifiable-Rule & Agent & Yes
      & Verifiable security/safety requirements governing agent behavior.
      & \textit{``If any portion of the flight has already been flown, the agent cannot help''}~\cite{yao2024tau}, a verifiable requirement for an airline-ticket agent. \\
    Task-Specific & Task & Yes
      & Verifiable security/safety requirements apply only to a specific task.
      & For a web agent over arbitrary webpages, \textit{``When clicking the Create group button, ask permission''}~\cite{levy2026stwebagentbench} applies to a GitLab group-creation task but not, e.g., to an online-shopping task. \\
    \bottomrule
  \end{tabular}
\end{table*}

The query returned 553 candidate papers. Manual inspection revealed that many were irrelevant, concerning reinforcement learning or robotics, where ``agent'' and ``policy'' carry different meanings. We therefore added two exclusion criteria: (f) the paper is cross-listed in cs.RO; and (g) title or abstract contains any whole word from \{\textit{robot, self-driving, embodied, reinforcement}\}. After the exclusions, 413 papers remained.

\textbf{Filtering Methods.} We assessed whether each of the 413 papers proposes benchmarks. One author manually annotated a random sample of 100 papers, while GPT-5-nano annotated all 413; both the human and the LLM saw only the paper title and abstract. On the 100-paper sample, human and LLM labels reached a Cohen's kappa of 0.88 (95\% CI $[0.76, 0.99]$), with LLM precision 0.96 and recall 0.99 against the human labels. Given this high agreement, we adopted the LLM's labels, filtering out 112 non-benchmark papers, leaving 301.

We manually inspected all 301 papers. We excluded 11 for not proposing a benchmark, 135 for not targeting tool-calling LLM-based agents, and 52 for not evaluating security or safety aspects. We further excluded 23 as out-of-scope: benchmarks assessing agents' ability to perform security- or safety-related tasks rather than evaluating the security or safety of the agents themselves; extensions of already included benchmarks that add only data or metrics without altering security or safety policies; and evaluations of whether an agent threat is plausible rather than whether the agent is secure or safe. As in \figurename~\ref{fig:prisma_chart}, 80 papers remained; details are available in the paper artifact.

\subsubsection{Classifying Benchmarks}
For each of the 80 benchmark, we analyze its security or safety policy, defined as the set of security and safety requirements the benchmark specifies.
Using forced-choice coding, we classified benchmarks by (1) the target agent type (general-purpose vs. domain-specific, Sec.~\ref{sec:related_works_agents}) and (2) policy specificity, characterized by scope and verifiability (\tablename~\ref{tab:policy_types}).
The focus here is \textit{verifiability}: given a policy and an agent's actions, whether one can prove (verify) at the level that the actions satisfy the policy~\cite{ISOIECIEEE29148}. 
Verifiability is an established quality attribute of well-formed requirements in requirements engineering~\cite{ISOIECIEEE29148} and relates to research on ambiguity in natural-language requirements~\cite{Berry2003Ambiguity}. In our corpus, the issue is more often \textit{omission} than \textit{imprecision}: requirements either leave the operative term (e.g., \textit{unsafe} or \textit{private}) unexplained, or benchmarks state no requirement at all, conveying expected behavior only through evaluation data. The problem is often not that a requirement could be read two ways (ambiguity), but that too little is stated for verification.

We do not attempt to define verifiability thresholds precisely, and we do not believe doing so would be productive: like specificity and ambiguity in software requirements, verifiability lies on a spectrum, and any threshold would be arbitrary. This stance is shared by work on automated requirements-quality checkers, where whether a flagged issue is a genuine problem is a contextual judgment~\cite{FEMMER2017190, 8254303}. We use four categories for a pragmatic partition: \textit{No Policy}, \textit{Goal-Setting}, \textit{Verifiable-Rule}, and \textit{Task-Specific}, defined and illustrated in \tablename~\ref{tab:policy_types}. In practice, the distinction is usually clear, and all authors discussed the rare borderline cases; we release all data for readers to confirm. We classified five benchmarks with insufficient information as \textit{unclear}.
\textit{Goal-Setting} policies fall into two patterns: (1) using \textit{safety} or \textit{security} without explanation, e.g., \textit{``If user requests anything unsafe, do not fulfill''}~\cite{zhou2026safeproevaluatingsafetyprofessionallevel}; and (2) leaving key terms unexplained, typically adjectives constraining information (e.g., \textit{private}, \textit{sensitive}) or actions (e.g., \textit{necessary}, \textit{legitimate}), e.g., \textit{``If query asks private information, decline to answer''}~\cite{huang2026crmarenapro}.
% and \textit{``Tool execution permitted ONLY when `absolutely necessary'\,''}~\cite{qiao2026agenttoolsorchestrationleaks}.

\subsubsection{Threats to Validity}
First, we use benchmark policies as a proxy for the security and safety requirements that deployed agents should satisfy. Benchmarks reflect their developers' interests and may characterize researchers' attention more than deployment concerns.
Second, despite arXiv's broad coverage of ML-related papers\footnote{For each of NeurIPS, ICML, ICLR, ACL, EMNLP, and NAACL, we sampled 100 papers accepted in 2025 and searched arXiv for exact title matches. Hit rates were 71\%, 77\%, 78\%, 74\%, 71\%, and 74\%. Manual inspection showed a lot of misses were slight title variants of papers already on arXiv, indicating arXiv indexes the vast majority of ML papers.} and our broad selection criteria, we may have missed benchmarks. Because this section reports policy-type distributions while later sections focus on \textit{Verifiable-Rule} benchmarks, we performed an informal snowballing search, tracking citations to and from known papers, to confirm that our \textit{Verifiable-Rule} set is complete. Among 61 candidates, we identified no new \textit{Verifiable-Rule} benchmarks, suggesting this limitation does not alter our conclusions.
Finally, our benchmark classifications are categorical, though domain and policy specificity lie on a spectrum and require judgment. We applied labels consistently and discussed uncertain cases jointly. All data are available for external validation.

\subsection{Results}\label{sec:part_result}

\begin{table}[t]
 \caption{Distribution of Benchmarks by Agent Domain and Policy Specificity, Computed over $n=80$}
    \label{tab:domain_policy_stat}
    \centering
    \begin{tabular}{lccc}
      \toprule
    & \multicolumn{2}{c}{\textbf{Agent Domain}} & \\
    \cmidrule(lr){2-3}
    \textbf{Policy Specificity} & \textbf{Gen.-Purp.} & \textbf{Dom.-Spec.} & \textbf{Total} \\
      \midrule
      No Policy     & 38 (47.5\%) & 11 (13.8\%) & 49 (61.3\%) \\
      Goal-Setting  & 14 (17.5\%) & 5 (6.3\%)   & 19 (23.8\%) \\
      Verifiable-Rule& 0 (0.0\%)   & 5 (6.3\%)   & 5 (6.3\%) \\
      Task-Specific & 1 (1.3\%)   & 1 (1.3\%)   & 2 (2.5\%) \\
      Unclear       & 2 (2.5\%)   & 3 (3.8\%)   & 5 (6.3\%) \\ \midrule
      \textbf{Total} & 55 (68.8\%) & 25 (31.3\%) & 80 (100.0\%) \\
      \bottomrule
    \end{tabular}
\end{table}

We identified 80 benchmarks for AI agent security and safety. \tablename~\ref{tab:domain_policy_stat} summarizes them along two dimensions: agent type (general-purpose vs.\ domain-specific) and the specificity of their security and safety policies.

\textbf{Security and safety requirements in 61\% of agent benchmarks are left implicit.}
These benchmarks provide no explicit instruction for the agent to behave safely or securely; their designers implicitly expect agents to prioritize security and safety norms over user instructions, even though this expectation is never communicated. For example, AgentHarm~\cite{andriushchenko2025agentharm} expects agents to refuse a user request to forge a passport, without instructing the agent that safety should override obedience.
In effect, designers expect agents to follow \textit{common-sense} security and safety norms, yet what counts as common sense varies across benchmarks and is rarely articulated, leaving it to be inferred from evaluation criteria.

\textbf{Security and safety encompass broad themes and are interpreted differently across benchmarks.}
Conceptions of security and safety vary widely across benchmarks (Sec.~\ref{sec:definingsafety}), spanning themes such as privacy, authorization, robustness to attacks, fairness, truthfulness, bodily harm, and policy compliance. Even the same theme is often interpreted in different, sometimes conflicting ways. For user privacy, for example, MobileSafetyBench~\cite{Lee_Hahm_Choi_Knox_Lee_2026} expects an agent to always request user permission before sharing private information. MPCI~\cite{wang2026mpcibenchbenchmarkmultimodalpairwise}, in contrast, evaluates whether an agent shares confidential information in ``suitable'' social contexts and refuses in inappropriate ones, without providing explicit instructions.

\textbf{Verifiable policies (\textit{Verifiable-Rule} and \textit{Task-Specific}) are rare (9\%).}
In non-verifiable policies, what counts as dangerous, illegal, or private is often unclear or context-dependent. Only 9\% of benchmarks specify plausibly verifiable requirements. For example, CAR-bench~\cite{kirmayr2026carbenchevaluatingconsistencylimitawareness} states: \textit{``if tool description starts with REQUIRES\_CONFIRMATION, before calling \dots\ must list the tool parameter and action details and obtain explicit expressive user confirmation\dots''} This rule is verifiable because both the antecedent (\textit{REQUIRES\_CONFIRMATION} prefix) and the consequent (explicit content for user confirmation) are checkable from the agent's trace.

\subsection{Discussion}\label{sec:part1_discussion}

\textbf{\textit{Verifiable-Rule} and \textit{Task-Specific} policies are essential for security and safety assurances in commercial deployments.}
When policies are implicit or unverifiable, practitioners can neither implement reliable guardrails nor audit agent behavior against a standard. \textit{Common-sense} expectations cannot fill this gap: many hold only within a community or are contested. For example, LPS-Bench~\cite{chen2026lpsbenchbenchmarkingsafetyawareness} considers agent that \textit{``adheres too rigidly to instructions while ignoring implicit intent''} as unsafe, which we find contestable: in high-stakes scenarios such as medical record updates, following explicit instructions may be preferable to inferring intent.
In commercial deployments, violating a requirement can be costly, and relying on the underlying model to interpret vague goals (e.g., what counts as \textit{necessary} or \textit{private}) shifts this risk onto an inductive process that cannot be reasoned about. \textit{Verifiable-Rule} and \textit{Task-Specific} policies remove this dependence by stating, in verifiable form, which actions the agent must or must not take, providing a basis for enforcement guardrails.

\textbf{\textit{Verifiable-Rule} policies are preferable to \textit{Task-Specific} policies.}
Both contain verifiable requirements, but \textit{Verifiable-Rule} policies state agent-level requirements that generalize across tasks, while \textit{Task-Specific} policies apply only to one specific task and require generating and updating policies per use case. In practice, this means either delegating policy generation to a model, which reintroduces the reliability concerns that motivated verifiable policies in the first place, or asking the user to articulate the policy for every input in a verifiable form (e.g. formal notation). Both options are problematic: the former undermines reliability, while the latter is impractical in most deployments considering usability and cost.

\textbf{Domain-specific agents are more feasible for enumerating \textit{Verifiable-Rule} policies.}
Although verifiable requirements are desirable, articulating them for general-purpose agents is hard; their broad use cases make \textit{common-sense} expectations seem more scalable. Consistent with this, all \textit{Verifiable-Rule} benchmarks we found target domain-specific agents (\tablename~\ref{tab:domain_policy_stat}): inputs may vary, but the task scope is defined and tools are limited, so verifiable requirements can be enumerated in advance.
For example, an airline-ticket agent with access only to airline-database tools can be given the requirement \textit{``If flights are flown, agent cannot cancel and refund''}~\cite{barres2025tau}. For general-purpose agents with open-ended tasks and broad tool access, comprehensively specifying verifiable requirements is far more difficult.
Yet 20 of 25 domain-specific benchmarks still do not specify \textit{Verifiable-Rule} policies (\tablename~\ref{tab:domain_policy_stat}), a missed opportunity that we argue should be addressed.
We believe many settings exist where domain-specific agents can be deployed safely and securely while providing practical value, whereas general-purpose agents remain too risky even when broadly aligned with \textit{common-sense} expectations. Domain-specific agents with enforceable security and safety are therefore a promising direction for commercial deployment.

\section{Study 2: Symbolic Enforceability of Security and Safety Requirements in Agent Benchmarks}\label{sec:part2_analyzing_policies} 

Study 1 found that most agent benchmarks do not state verifiable security and safety requirements. Study 2 examines the few that do, since they are the only benchmarks that admit security and safety guarantees needed in risk-averse commercial settings. We ask: \textit{among verifiable security and safety requirements in agent benchmarks, which can be symbolically enforced, and how? Why are others not enforceable?}

This restricts Study 2 to the seven benchmarks with verifiable security and safety requirements identified in Sec.~\ref{sec:part1_collecting_policies} (5 \textit{Verifiable-Rule} and 2 \textit{Task-Specific}). This is not a sampling decision: these seven are the only benchmarks that can answer our research questions on symbolic enforceability.

%since the others state no verifiable requirements in the first place. %Our results thus reflect the requirements that this part of the research community tries to enforce with no statistical generalization intended.

\subsection{Research Method}\label{sec:part2_research_method}

\begin{table*}[t]
 \caption{Symbolic Guardrails and Illustrative Examples }
  \label{tab:symbolic_guardrail_list}
    \centering
  \begin{tabular}{p{3.3cm}l}
    \toprule
    \textbf{Symbolic Guardrail} & \textbf{Example$^{\mathrm{a}}$} \\
    \midrule
    API Validation & Before invoking \verb|cancel_ticket(user, ticket)|, verify that \verb|user == ticket.user|.
    \\ 
    Schema Constraint~\cite{scholak-etal-2021-picard} & Reject LLM output if it is neither a tool invocation matching the airline API schema nor a message to the user. \\
    Temporal Logic~\cite{wang2025agentspeccustomizableruntimeenforcement} & Block all other tools until \verb|authenticate_user| completes successfully. \\
    Information Flow~\cite{costa2025securingaiagentsinformationflow} & Prevent information about other passengers from reaching the agent.\\
    User Confirmation & Before \verb|cancel_ticket|, require rule-based user confirmation rather than LLM-initiated confirmation.\\
    Response Template & After \verb|cancel_ticket|, display a predefined cancellation summary rather than an LLM-generated response. \\
    \bottomrule
    \multicolumn{2}{p{17cm}}{$^{\mathrm{a}}$Illustrative examples are based on an airline ticket agent, inspired by $\tau^2$-Bench~\cite{barres2025tau}. The agent assists users with managing their flight bookings and interacts with internal databases by invoking tools such as \texttt{get\_flight\_info}, \texttt{get\_user\_info}, and \texttt{cancel\_ticket}.}
  \end{tabular}
\end{table*}

\subsubsection{Selecting Benchmarks}

Among the seven benchmarks with verifiable security and safety requirements, we exclude the two \textit{Task-Specific} ones, where guardrail applicability would be entirely input-dependent. Of the five \textit{Verifiable-Rule} benchmarks, four target customer-service agents. To avoid overfitting to one domain, we select $\tau^2$-Bench\cite{barres2025tau} (the most widely used of the four) and CAR-bench\cite{kirmayr2026carbenchevaluatingconsistencylimitawareness} (the non-customer-service benchmark, for in-car voice assistants). Both state policies in natural language; we treat each sentence as a candidate requirement, yielding 120 in $\tau^2$-Bench and 18 in CAR-bench.

To broaden domain coverage beyond these two, we considered synthesizing plausible \textit{Verifiable-Rule} policies for additional domain-specific benchmarks.
We initially planned to construct policies for the 20 domain-specific benchmarks without \textit{Verifiable-Rule} policies, so policies reflect what the benchmark evaluates. This proved difficult: although evaluation methods implicitly encode security and safety expectations, recovering consistent verifiable requirements would have demanded extensive data review and modification, as task conflicts were common. For example, in CRMArenaPro~\cite{huang2026crmarenapro}, one task expects the agent to reject \textit{``Considering recent discussions, should this lead be considered qualified?''} as a privacy violation, while another expects it to answer the near-identical query \textit{``After assessing recent discussions, should this lead be considered qualified?''} No single requirement aligns with both cases without modifying the benchmark labels.

As a result, we adopted an alternative solution, creating a synthetic policy for a domain-specific benchmark in a high-stakes domain that did not originally target security or safety, hence excluded from Study 1.
We selected MedAgentBench~\cite{jiang2025medagentbench}, which evaluates an electronic medical record (EMR) assistant.
To create a plausible policy without biasing it toward or against symbolic enforcement, we followed the procedure below.
First, we prompted GPT-5.2 to generate a security and safety policy from the EMR assistant's use case and tool schema, requesting verifiable rules with no mention of enforcement. This produced 50 requirements.
Second, we improved policy comprehensiveness through hazard analysis, specifically System-Theoretic Process Analysis (STPA)~\cite{Engineering_a_safer_world_Nancy2012}, which anticipates harms from diverse perspectives and derives corresponding requirements. Using an automated STPA tool from prior work~\cite{11030006} with GPT-5, we identified 5{,}138 candidate requirements. To keep scope manageable, we randomly sampled 4\% (205 entries), clustered them into 10 categories via K-Means over OpenAI \texttt{text-embedding-3-small} embeddings, and prompted an LLM to consolidate each cluster, yielding 77 candidate requirements; we randomly sampled 20 for further analysis. Because these 20 often contained compound logic, we decomposed them into 43 atomic requirements; after removing 5 duplicates, 38 remained. Combining the 50 initial requirements and the 38 from STPA produced a final policy of 88 requirements. All key steps were automated to avoid author bias; full details are in the paper artifact.

\subsubsection{Classifying Requirements Enforceability and Matching Symbolic Guardrails}\label{sec:scope_of_sym_guardrail}

For each requirement, we manually classify whether it can be plausibly enforced symbolically, that is, through deductive reasoning in program code. We operationalize this by checking whether any of six symbolic guardrails, explained in \tablename~\ref{tab:symbolic_guardrail_list}, could enforce it: API validation, schema constraints, information flow\footnote{Strict enforcement of information-flow properties is challenging given implicit flows and environment interactions, though various approaches exist~\cite{sabelfeld2003language}; these nuances are beyond our scope.}, temporal logic, user confirmation, and response templates. 
We assembled this catalog in two ways. Top-down, we identified methods already used in prior work to secure AI agents (Sec.~\ref{sec:related_works_guardrails}), such as information-flow control~\cite{costa2025securingaiagentsinformationflow}. Bottom-up, we identified strategies not usually discussed in the academic literature but well-suited to the requirements these benchmarks evaluate.

For each candidate requirement in the three benchmarks' policies, we manually assign one of three labels:

\begin{itemize}[leftmargin=*]
    \item \textbf{Out of scope:} (1) provides information rather than stating a requirement (e.g., ``User's profile contains their email''); (2) specifies a system-level requirement that cannot be fulfilled through agent behavior (e.g., ``No PII should be stored persistently''); or (3) was hallucinated by the LLM and is infeasible given the available tools (MedAgentBench only).
    \item \textbf{Enforceable symbolically:} can be enforced by one or more symbolic guardrails; we also record which guardrails.
    \item \textbf{Likely not enforceable symbolically:} cannot be enforced by any combination of the considered guardrails.
\end{itemize}

In most cases, classification was straightforward. To validate borderline labels, all authors discussed them and reached consensus; we share our classifications for readers to confirm. For each requirement labeled enforceable, one author implemented the corresponding guardrail and verified its behavior against the benchmark data; these implementations are reused in Study 3 (see Sec.~\ref{sec:part3_experiments} for details). All authors also discussed requirements classified as not enforceable, exploring common reasons through a card-sorting-style grouping and reflection.

\subsubsection{Threats to Validity}
Our analysis is limited by the small number of \textit{Verifiable-Rule} benchmarks, covering only two domains. The MedAgentBench policy extends coverage but is LLM-generated rather than human-written and may not fully reflect requirements practitioners would impose in deployment. Hazard analysis helps identify security and safety concerns comprehensively, but we followed an automated process without expert judgment. As before, benchmark requirements often shape what researchers focus on and may not match the concerns of commercial settings. Our results should therefore be interpreted as a step toward understanding what enforcement is possible for \textit{plausible} domain-specific policies, not as a comprehensive reflection of commercial practice.

Our guardrail catalog may also be incomplete, and our judgment of symbolic enforceability also depends on our understanding of what symbolic guardrails can achieve and our ability to implement them. Therefore, the enforceable rate should be read as a conservative lower bound, which does not undermine our argument.

\subsection{Which requirements are enforceable with symbolic guardrails, and how?}\label{sec:rq2}

\begin{figure}[t]
\centering
  \includegraphics[width=\linewidth]{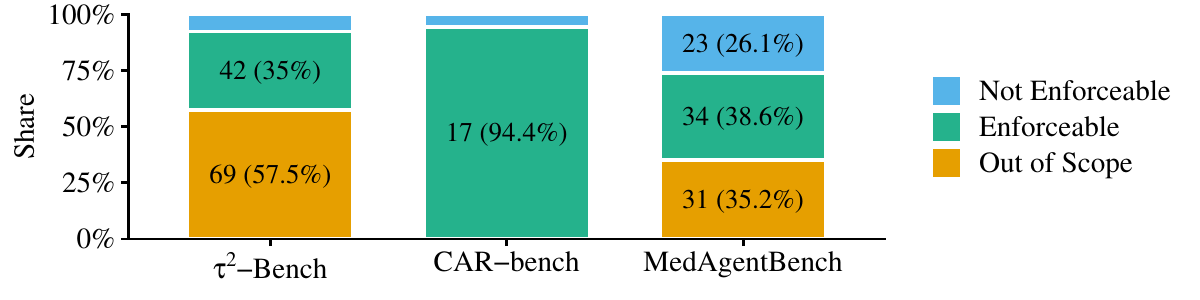}
 \caption{Distribution of security and safety requirements enforceability across the three benchmarks.}
  \label{fig:enforceability}
\end{figure}

\begin{figure}[t]
\centering
  \includegraphics[width=\linewidth]{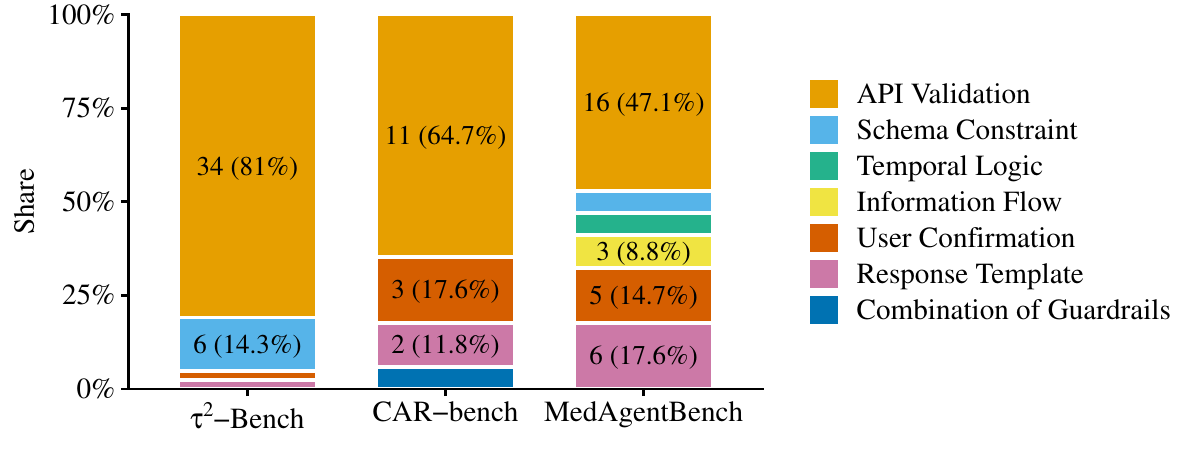}
 \caption{Distribution of applicable symbolic guardrails for enforceable requirements across the three benchmarks.}
  \label{fig:symbolic_guardrails}
\end{figure}

\subsubsection{Results}\label{sec:rq2_result}

% \figurename~\ref{fig:enforceability} shows enforceability across the three benchmarks; \figurename~\ref{fig:symbolic_guardrails} breaks down which guardrails apply.

\textbf{Across the three benchmarks, 74\% of security and safety requirements are enforceable by symbolic guardrails.} As shown in \figurename~\ref{fig:enforceability}, after excluding out-of-scope requirements, symbolic guardrails enforce 42/51 in $\tau^2$-Bench, 17/18 in CAR-bench, and 34/57 in MedAgentBench.

\textbf{Simple guardrails such as API validation often suffice.} As shown in \figurename~\ref{fig:symbolic_guardrails}, API validation alone covers 81\%, 65\%, and 47\% of enforceable requirements in $\tau^2$-Bench, CAR-bench, and MedAgentBench, respectively. Adding other simple methods (schema constraints, user confirmation, and response templates) brings coverage to 95\% of all enforceable requirements. Only five requirements (all in MedAgentBench) require temporal logic or information-flow control, which are relatively more costly to implement.

\subsubsection{Discussion}\label{sec:rq2_discussion}

\textbf{Symbolic guardrails are effective and inexpensive for many security and safety requirements in the analyzed agents.}
Many security and safety requirements in identified policies can be enforced with symbolic guardrails, and in most cases are inexpensive in both engineering effort and runtime cost, because the requirements can be handled through simple API validation, schema enforcement, or hard-coded user confirmation and response templates. These mechanisms are rarely discussed in research on AI agent security and safety. More sophisticated mechanisms explored in recent work, such as information-flow tracking and temporal logic, require greater engineering effort and are rarely needed (5 of 93, 5\%) for the enforceable requirements in the three analyzed benchmarks. Even the few we identified that require these methods are simple, for example, ``Block all other tools until \verb|authenticate_user| completes successfully.''
We argue that domain-specific agents offer many \textit{low-hanging fruit} opportunities: lightweight checks that could prevent many agent errors. Surprisingly, benchmark implementations often do not enforce even basic rules in the tools themselves, for example, that an agent cannot cancel a flight already flown, and instead rely on the agent model or additional neural guardrails. This design violates basic software engineering and security principles such as least privilege and complete mediation~\cite{viega2001building}.

\textbf{Symbolic guardrails may simplify agent prompts and improve instruction following.}
Once enforced symbolically, requirements can be removed from the agent's prompt, reducing context size and token costs. Because LLMs struggle to follow instructions when too many are provided~\cite{yang2025promptsdontsayunderstanding}, a shorter policy may also improve compliance with the remaining instructions. At the same time, keeping some enforced requirements in the policy may still be useful, so the agent can take the right action upfront rather than receive a guardrail error. This trade-off is an implementation choice beyond our scope, and one that, to our knowledge, has received little attention in the agent security and safety literature.

% \textbf{Some agent security or safety benchmarks may be less informative.}
% As discussed in Sec.~\ref{sec:analyzing_safety_and_security_policies}, we were unable to formulate Verifiable Rule policies for many existing agent security and safety benchmarks. In the absence of Verifiable Rule policies, the data construction of these benchmarks often depends on model-generated outputs followed by human filtering. When the underlying policies are ambiguous, this process can introduce conflicting examples into the data. For instance, in CRMArenaPro~\cite{huang2026crmarenapro}, one task requires the agent to reject the query ``Considering the recent discussions with the lead, should this lead be considered qualified?'' because it violates privacy policy, while another task expects the agent to answer the closely similar query ``After assessing the lead's recent discussions, should this lead be considered qualified?'' These inconsistencies suggest that these benchmarks may provide a weaker signal about true agent security or safety performance.

\subsection{Which requirements are not enforceable with symbolic guardrails, and what alternative approaches are available?}\label{sec:rq3}

\subsubsection{Results}\label{sec:rq3_result}

\textbf{We identified four recurring themes among the not enforceable requirements in the three benchmarks}, described below; mitigations are discussed in Sec.~\ref{sec:rq3_discussion}.

\textbf{Persona and interaction-style requirements} specify how the agent should communicate, such as using specific language or tone. For example, a medical assistant may be required to remain neutral and avoid offering medical judgment~\cite{jiang2025medagentbench}.

\textbf{``No hallucination'' requirements} expect the agent to avoid generating unsupported or fabricated information. For example, $\tau^2$-Bench specifies that the agent ``should not provide any information not provided by the user or available tools''~\cite{barres2025tau}.

\textbf{Procedure-following requirements} specify that the agent must follow a predefined sequence of steps. For example, in $\tau^2$-Bench, an agent helping with airline booking is expected to first obtain user details and then trip details~\cite{barres2025tau}.

\textbf{Common-sense reasoning requirements} arise even within \textit{Verifiable-Rule} policies. These requirements leave substantial room for interpretation and therefore depend on \textit{common-sense} reasoning. For instance, in the flight-assistant setting of $\tau^2$-Bench, a policy such as ``Do not proactively offer compensation unless the user explicitly asks for it'' still requires the model to interpret what counts as an explicit request~\cite{barres2025tau}.

\subsubsection{Discussion}\label{sec:rq3_discussion}
\textbf{Symbolic guardrails cannot enforce all requirements; some still require neural guardrails.} The four themes above are not enforceable symbolically, even within \textit{Verifiable-Rule} policies. In such cases, neural guardrails can help: for example, an LLM judge may detect hallucinations while symbolic guardrails cannot.

\textbf{Some non-enforceable requirements become enforceable through stronger or weaker reformulations; whether to do so is an engineering decision.}
For example, a procedure-following requirement to collect user information before trip details when booking a flight could be enforced via a sub-agent design that splits the workflow into sequential stages, which is a stronger reformulation requiring architectural change. Conversely, the requirement \textit{``Do not proactively offer compensation unless the user explicitly asks for it''} could be replaced with a weaker but more precise rule such as \textit{``Block compensation tools until the user's response exactly matches the word `compensation'\,''}, making it enforceable.
In commercial deployments, this is a trade-off between implementation cost and residual risk. Enforcing requirements symbolically whenever feasible reduces both the attack surface and the potential for hazards, letting developers reserve more expensive neural guardrails with residual risk for the remaining requirements.

\section{Study 3: Benchmarking Security, Safety, and Utility of Agents with Symbolic Guardrails}\label{sec:part3_experiments}

Symbolic guardrails can enforce certain security and safety requirements, but a concern is that they over-constrain agents and reduce utility (their ability to complete tasks successfully)~\cite{tan2025equilibraterlhfbalancinghelpfulnesssafety, dong2025safeguarding}. 
We ask: \textit{how do symbolic guardrails affect agent security, safety, and utility?}
Study 3 empirically measures this across the three benchmarks from Sec.~\ref{sec:part2_analyzing_policies}.

\subsection{Research Method}\label{sec:part3_experiment_method}

We run each benchmark with and without symbolic guardrails (independent variable) and measure policy violations (agent actions that violate symbolically-enforceable security and safety requirements) and utility (dependent variables). We reuse the three benchmarks from Study 2, $\tau^2$-Bench (airline)~\cite{barres2025tau}, CAR-bench~\cite{kirmayr2026carbenchevaluatingconsistencylimitawareness}, and MedAgentBench~\cite{jiang2025medagentbench}, along with the guardrails implemented there, detailed below.

\subsubsection{Experiment Infrastructure}

We implement a tool-calling agent following $\tau^2$-Bench~\cite{barres2025tau}, with GPT-4o and GPT-5 as backbone LLMs and policies in system prompts. Across all conditions, tools are exposed via an MCP server~\cite{MCP_2024}.

%Following prior work~\cite{yao2024tau}, the agent interacts with the MCP server under the constraint that, at each step, it may either make a single tool call and receive the tool call results or respond directly to the user. 
%The backbone model is constrained to generate tool calls that must strictly conform to the provided JSON Schema, using ``strict'' tool mode in OpenAI's API. 

Because these benchmarks require multi-turn user-agent interactions, we simulate user responses with an LLM, in line with $\tau^2$-Bench and CAR-bench. The user-simulation LLM is given the relevant context but has no access to the MCP tools.

\subsubsection{Tool and Symbolic Guardrail Implementation}\label{sec:tooland symguardrail implementation}

For $\tau^2$-Bench and CAR-bench, we use the tool implementations released with each benchmark as the \textit{baseline} condition, wrapping them with symbolic guardrails for all enforceable requirements to form the \textit{guardrail} condition (\figurename~\ref{fig:symbolic_guardrails}). 
% Agent-side guardrails, such as user confirmation, are implemented as agent harnesses and triggered via MCP tool metadata.

For MedAgentBench, we implement three tool conditions: (1) the \textit{raw} condition uses an MCP server exposing generic GET and POST tools, mirroring the original benchmark design in which the agent constructs HTTP requests autonomously; (2) the \textit{baseline} condition uses an MCP server with eight tools, one per HTTP endpoint in the original benchmark; and (3) the \textit{guardrail} condition wraps the eight-tool baseline with symbolic guardrails for 23 enforceable requirements, drawn from the 34 classified as enforceable in Study 2. Of the remaining 11, one was already implemented in the benchmark; the other ten require substantial medical domain knowledge an expert could enumerate but was not available to us, such as appropriate dosage units for different medications.

Across all three benchmarks, \textit{baseline} and \textit{guardrail} share tool schemas, but \textit{guardrail} adds enforcement parameters in 6 of 16 $\tau^2$-Bench tools and 6 of 8 MedAgentBench tools. For example, $\tau^2$-Bench's \verb|cancel_ticket| takes only \verb|ticket_id| in \textit{baseline} and adds \verb|user_id| in \textit{guardrail} to verify ownership via \verb|user_id == ticket.user|. CAR-bench needs no additional parameters.
 %Aside from such additional guardrail-related arguments, the tool schema remains unchanged.

\subsubsection{Datasets}

For $\tau^2$-Bench, we use $\tau^2$-Bench-Verified (airline)~\cite{cuadron2025sabersmallactionsbig}, a cleaned version that fixes inconsistencies in the original benchmark without altering the policy, prompts, or tools. It contains 50 tasks in which AI agents assist customers with flight reservations; some customers attempt policy violations, e.g., requesting a refund for a non-refundable flight.

For CAR-bench, we use the 100 data entries from the ``Base'' category, in which the agent acts as an in-car voice assistant that helps users with navigation and vehicle operations such as checking the weather and adjusting fog lights.

MedAgentBench evaluates whether an agent can assist with electronic medical record (EMR) tasks such as checking patient records and ordering medications. It contains 300 tasks, none probing for policy violations. Because our policy requires patient authorization, which is not considered in the original, we augment the dataset by providing the \textit{user} (simulated by LLMs) with the patient information needed for authorization.

To supplement our analysis, we construct an \textit{adversarial dataset} for MedAgentBench in which the \textit{user} tries to manipulate the agent into violating a requirement. Following prior practices~\cite{huang2026crmarenapro, levy2026stwebagentbench, chen2026lpsbenchbenchmarkingsafetyawareness}, we identified four task categories from the original benchmark and prompted an LLM to expand them into 17 scenarios. For each scenario-requirement pair, the LLM generated an adversarial task that seeks to violate the requirement within the scenario. Due to cost, we sampled 50 of the 391 tasks (17 scenarios $\times$ 23 requirements) for evaluation.

\subsubsection{Dependent variables}

We measure \textit{utility} as the first-try task success rate, using each benchmark's original metric: Pass\textasciicircum{}1 for $\tau^2$-Bench and CAR-bench, and Success Rate (SR) for MedAgentBench. We do not measure utility on the adversarial dataset, as task completion is not expected.

For \textit{security and safety}, only CAR-bench measures policy violations among the three: its $r_{policy}$ metric evaluates whether each task follows the policy throughout, relying in part on a probabilistic LLM judge.
To evaluate \textit{security and safety} non-probabilistically, we restrict evaluation to symbolically enforceable requirements, the only ones that can be validated accurately and are addressed in this work. By construction, our symbolic guardrails prevent violations by rejecting all noncompliant tool calls; we therefore measure policy violations only under \textit{raw} and \textit{baseline}, where guardrails are absent.

To detect requirement violations, we also apply the symbolic guardrails in the \textit{raw} and \textit{baseline} conditions, but only to \textit{record} triggers rather than \textit{reject} tool calls. We then count the tasks that trigger at least one violation in each benchmark.

A complication arises when a guardrail adds extra tool parameters (Sec.~\ref{sec:tooland symguardrail implementation}), affecting 6 of 16 tools in $\tau^2$-Bench and 6 of 8 in MedAgentBench. Because modifying tool signatures could alter agent behavior, we adopt a \textit{replay-based evaluation procedure} to avoid biasing the agent: when the agent calls a \textit{baseline} tool that would require extra arguments under the \textit{guardrail} condition for policy violation checks, we pause execution, roll back to the state before the \textit{baseline} tool-call decision, and re-prompt the model with the extended \textit{guardrail} signature. If the replayed call uses the same tool name and original arguments and supplies the required extra argument, we use it to assess policy violations. After this assessment, we resume execution from the rollback point with the original \textit{baseline} call, so that the evaluation does not alter the agent's behavior. This yields the information needed for policy-violation checks, such as the user ID for \verb|cancel_ticket|, without altering the reasoning induced by the \textit{baseline} tool schema. If the agent cannot supply the additional argument during replay, we classify the execution as \textit{unsafe}, since the context is insufficient for a security or safety check. If it instead selects a different tool or changes the original arguments, we retry up to five times; if the call still cannot be reproduced, we label the outcome as \textit{unknown}.

Since CAR-bench requires no additional parameters under \textit{guardrail} condition, no tool calls receive an \textit{unknown} outcome, so we do not report its \textit{unknown} rate for policy violation.

For all dependent variables, we assess the significance of differences across conditions using the paired McNemar test.

\subsubsection{Threats to Validity}
Our evaluation covers only policy violations that are reliably detectable via symbolic guardrails and excludes those that are not. As is common in agent benchmarks, executions are expensive even at benchmark sizes of 50 to 300 tasks, with a single benchmark run costing approximately USD 80. This constrains the experimental conditions (e.g., model choices) and repetitions we can evaluate, so our findings reveal general trends, and the statistical tests may detect only large effects. Further work is needed to assess generalization beyond the three benchmarks. Finally, all experimental conditions include the full policy in the system prompt, even though many policy sentences are technically redundant under the \textit{guardrail} condition; future work could examine the trade-offs of removing this redundancy.

\subsection{Results}\label{sec:part3_result}
\begin{table}[t]
\caption{Experiment Results on Three Benchmarks}
\label{tab:study3_results}
\centering

\begin{subtable}{\linewidth}
\centering
\caption{$\tau^2$-Bench}
\label{tab:tau2}
\begin{tabular}{cccccc}
\toprule
 & & \multicolumn{3}{c}{\textbf{Security \& Safety}} & \textbf{Utility}\\
\cmidrule(lr){3-5}\cmidrule(lr){6-6}
\textbf{Model} & \textbf{Condition} & \textbf{Unsafe$\downarrow$} & \textbf{Unknown} & \textbf{Safe$\uparrow$} & \textbf{Pass\textasciicircum1$\uparrow$}\\
\midrule
GPT-4o & \textit{baseline}  & 52.0\% & 0.0\%  & 48.0\%  & 0.36 \\
       & \textit{guardrail} & 0.0\%  & 0.0\%  & 100.0\% & 0.48 \\
GPT-5  & \textit{baseline}  & 20.0\% & 10.0\% & 70.0\%  & 0.68 \\
       & \textit{guardrail} & 0.0\%  & 0.0\%  & 100.0\% & 0.70 \\
\bottomrule
\end{tabular}
\end{subtable}

\vspace{0.8em}

\begin{subtable}{\linewidth}
\centering
\caption{CAR-bench (GPT-5)}
\label{tab:CAR-bench}
\begin{tabular}{ccccc}
\toprule
 & \multicolumn{3}{c}{\textbf{Security \& Safety}} & \textbf{Utility}\\
\cmidrule(lr){2-4}\cmidrule(lr){5-5}
\textbf{Condition} & \textbf{Unsafe$\downarrow$} & \textbf{Safe$\uparrow$} & \textbf{$r_{policy}\uparrow$} & \textbf{Pass\textasciicircum1$\uparrow$}\\
\midrule
\textit{baseline}  & 21.0\% & 79.0\%  & 0.83 & 0.59 \\
\textit{guardrail} & 0.0\%  & 100.0\% & 0.97 & 0.72 \\
\bottomrule
\end{tabular}
\end{subtable}

\vspace{0.8em}

\begin{subtable}{\linewidth}
\centering
\caption{MedAgentBench (GPT-5)}
\label{tab:medagentbench}
\begin{tabular}{cccccc}
\toprule
 & & \multicolumn{3}{c}{\textbf{Security \& Safety}} & \textbf{Utility}\\
\cmidrule(lr){3-5}\cmidrule(lr){6-6}
\textbf{Data} & \textbf{Condition} & \textbf{Unsafe$\downarrow$} & \textbf{Unknown} & \textbf{Safe$\uparrow$} & \textbf{SR$\uparrow$}\\
\midrule
Original    & \textit{raw}       & 39.0\% & 9.7\% & 51.3\%  & 0.64 \\
            & \textit{baseline}  & 23.0\% & 0.3\% & 76.7\%  & 0.59 \\
            & \textit{guardrail} & 0.0\%  & 0.0\% & 100.0\% & 0.67 \\
Adversarial & \textit{raw}       & 78.0\% & 4.0\% & 18.0\%  & -- \\
            & \textit{baseline}  & 62.0\% & 4.0\% & 34.0\%  & -- \\
            & \textit{guardrail} & 0.0\%  & 0.0\% & 100.0\% & -- \\
\bottomrule
\end{tabular}
\end{subtable}

\end{table}

\textbf{AI agents without guardrails violate requirements that symbolic guardrails can enforce.}
As shown in \tablename~\ref{tab:study3_results}, policy violations are common across all settings without symbolic guardrails: 20\% to 78\% of task executions violate symbolically enforceable requirements. Violations are more frequent on adversarial tasks (\tablename~\ref{tab:medagentbench}) and for weaker models (compare GPT-4o and GPT-5 in \tablename~\ref{tab:tau2}), but occur under every condition without guardrails. Under the \textit{guardrail} condition, such violations are prevented by construction, and the difference is statistically significant ($p<0.01$ in all cases). AR-bench's $r_{policy}$ metric (\tablename~\ref{tab:CAR-bench}), evaluated in part by an LLM judge, shows the same pattern ($p<0.01$).

\textbf{Symbolic guardrails do not sacrifice agent utility.} As shown in \tablename~\ref{tab:study3_results}, utility increases under the \textit{guardrail} condition across all datasets, though several improvements are not statistically significant ($p=0.18$ for GPT-4o and $p=1.00$ for GPT-5 on $\tau^2$-Bench; $p<0.01$ on CAR-bench; $p=0.27$ vs. \textit{raw} and $p<0.01$ vs. \textit{baseline} on MedAgentBench). Overall, the results suggest that symbolic enforcement is unlikely to harm utility and may improve it.

% \textbf{Symbolic guardrails reduce unsafe tool-call attempts by the model. }  As shown in Tables~\ref{tab:tau_2_tool_call}--\ref{tab:medagentbench_adversarial_tool_call}, comparing the Baseline and Guardrail tool sets, symbolic guardrails reduce both the absolute number and the proportion of tool calls initiated by the agent that are unsafe (the sum of guarded and unsafe tool calls), regardless of whether those calls are ultimately rejected by the guardrail.

\subsection{Discussion}\label{sec:part3_discussion}

\textbf{Relying on models alone to enforce security and safety requirements is risky.} Even without adversarial users, agents frequently violate security and safety requirements stated explicitly in the system prompt. Stronger models make fewer mistakes, and future models with larger context windows and better instruction following may reduce these errors further, but easily preventable failures still create unnecessary risk. Dedicated neural guardrails can likely reduce many of these errors, but add non-trivial runtime cost and still leave residual risk. In adversarial settings, for example, prompt injection that causes a model to offer compensation improperly or prescribe the wrong medication, deploying such agents may become too risky to be feasible.
Although not all desirable security and safety properties can be enforced symbolically, practitioners in high-assurance commercial settings should assess whether the most important requirements can be specified and enforced this way, then use neural guardrails for the remaining critical requirements as part of a deliberate risk assessment.

\textbf{Security, safety, and utility are not necessarily a trade-off.}
Intuitively, guardrails seem to constrain agent flexibility, but our results suggest they actually help the model explore safe solutions more effectively. By inspecting the agent-interaction traces, we provide a possible explanation: when a symbolic guardrail blocks an unsafe action, it prevents the agent from terminating incorrectly and returns an error message explaining why the action failed, why it was insecure or unsafe, and which requirement it violated. The agent can then use this feedback to adjust, retry with a safer alternative, and often complete the task successfully.

\section{Conclusion}

We argue that symbolic guardrails are an \textit{overlooked but highly practical} mechanism for AI agent security and safety, especially in domain-specific, risk-averse commercial settings. Most existing benchmarks do not specify verifiable security or safety requirements; when they do, many can be enforced symbolically through simple methods. These guardrails eliminate a large class of security and safety violations without reducing agent utility. Symbolic guardrails are not a complete solution, as some requirements still depend on model-based judgment and neural guardrails, but using inductive methods for requirements that could instead be enforced symbolically introduces avoidable risk for limited benefit. We argue that broader use of symbolic guardrails is a promising path toward deploying domain-specific AI agents in high-stakes settings with stronger security and safety guarantees.

\section*{Acknowledgment}
This work was supported in part by the National Science Foundation (award 2206859) and an unrestricted gift from Google’s GARA.
We would also like to thank Chenyang Yang, the SSSG attendees, and the S3C2 Quarterly Meeting attendees for their valuable feedback on this work.

\newpage
\bibliography{references.bib}

@misc{barres2025tau,
     title={$\tau^2$-Bench: Evaluating Conversational Agents in a Dual-Control Environment}, 
      author={Victor Barres and Honghua Dong and Soham Ray and Xujie Si and Karthik Narasimhan},
      year={2025},
      eprint={2506.07982},
      archivePrefix={arXiv},
      primaryClass={cs.AI},
      url={https://arxiv.org/abs/2506.07982}, 
}

@article{jiang2025medagentbench,
  title={MedAgentBench: a virtual EHR environment to benchmark medical LLM agents},
  author={Jiang, Yixing and Black, Kameron C and Geng, Gloria and Park, Danny and Zou, James and Ng, Andrew Y and Chen, Jonathan H},
  journal={Nejm Ai},
  volume={2},
  number={9},
  pages={AIdbp2500144},
  year={2025},
  publisher={Massachusetts Medical Society}
}

@inproceedings{
yao2023react,
title={ReAct: Synergizing Reasoning and Acting in Language Models},
author={Shunyu Yao and Jeffrey Zhao and Dian Yu and Nan Du and Izhak Shafran and Karthik R Narasimhan and Yuan Cao},
booktitle={The Eleventh International Conference on Learning Representations },
year={2023},
no_url={https://openreview.net/forum?id=WE_vluYUL-X}
}

@inproceedings{schick2023toolformer,
 author = {Schick, Timo and Dwivedi-Yu, Jane and Dessi, Roberto and Raileanu, Roberta and Lomeli, Maria and Hambro, Eric and Zettlemoyer, Luke and Cancedda, Nicola and Scialom, Thomas},
 booktitle = {Advances in Neural Information Processing Systems},
 editor ={A. Oh and T. Naumann and A. Globerson and K. Saenko and M. Hardt and S. Levine},
 pages = {68539--68551},
 publisher = {Curran Associates, Inc.},
 title = {Toolformer: Language Models Can Teach Themselves to Use Tools},
 no_url = {https://proceedings.neurips.cc/paper_files/paper/2023/file/d842425e4bf79ba039352da0f658a906-Paper-Conference.pdf},
 volume = {36},
 year = {2023}
}

@inproceedings{Reflexion,
 author = {Shinn, Noah and Cassano, Federico and Gopinath, Ashwin and Narasimhan, Karthik and Yao, Shunyu},
 booktitle = {Advances in Neural Information Processing Systems},
 editor ={A. Oh and T. Naumann and A. Globerson and K. Saenko and M. Hardt and S. Levine},
 pages = {8634--8652},
 publisher = {Curran Associates, Inc.},
 title = {Reflexion: language agents with verbal reinforcement learning},
 no_url = {https://proceedings.neurips.cc/paper_files/paper/2023/file/1b44b878bb782e6954cd888628510e90-Paper-Conference.pdf},
 volume = {36},
 year = {2023}
}

@article{lewis2020retrieval,
  title={Retrieval-augmented generation for knowledge-intensive nlp tasks},
  author={Lewis, Patrick and Perez, Ethan and Piktus, Aleksandra and Petroni, Fabio and Karpukhin, Vladimir and Goyal, Naman and others},
  journal={Advances in neural information processing systems},
  volume={33},
  pages={9459--9474},
  year={2020}
}

@online{A2A_2025, 
    title={Announcing the agent2agent protocol (A2A)}, 
    url={https://developers.googleblog.com/en/a2a-a-new-era-of-agent-interoperability/}, 
    author={Surapaneni, Rao and Jha, Miku and Vakoc, Michael and Segal, Todd}, 
    year={2025}, 
    month={Apr},
    urldate={2026-03-02}
}

@online{MCP_2024, 
    title={Model Context Protocol}, 
    url={https://github.com/modelcontextprotocol}, 
    author={Anthropic}, 
    year={2024}, 
    urldate={2026-03-02}
}

@online{Chatgpt_agent, 
    title={ChatGPT agent}, 
    url={https://chatgpt.com/features/agent}, 
    author={OpenAI}, 
    year={2025}, 
    urldate={2026-03-02}
}

@online{copilot_windows, 
    title={Getting started with Copilot on Windows}, 
    url={https://support.microsoft.com/en-us/topic/getting-started-with-copilot-on-windows-1159c61f-86c3-4755-bf83-7fbff7e0982d}, 
    author={Microsoft}, 
    year={2023}, 
    urldate={2026-03-02}
}

@article{
huang2026crmarenapro,
title={{CRMA}rena-Pro: Holistic Assessment of {LLM} Agents Across Diverse Business Scenarios and Interactions},
author={Kung-Hsiang Huang and Akshara Prabhakar and Onkar Thorat and Divyansh Agarwal and Prafulla Kumar Choubey and Yixin Mao and Silvio Savarese and Caiming Xiong and Chien-Sheng Wu},
journal={Transactions on Machine Learning Research},
issn={2835-8856},
year={2026},
url={https://openreview.net/forum?id=EPlpe3Fx1x},
note={}
}

@article{ma2026safety,
  title={Safety at scale: A comprehensive survey of large model and agent safety},
  author={Ma, Xingjun and Gao, Yifeng and Wang, Yixu and Wang, Ruofan and Wang, Xin and Sun, Ye and others},
  journal={Foundations and Trends in Privacy and Security},
  volume={8},
  number={3-4},
  pages={1--240},
  year={2026},
  publisher={Emerald Publishing Limited}
}

@article{10.1145/3716628,
author = {Deng, Zehang and Guo, Yongjian and Han, Changzhou and Ma, Wanlun and Xiong, Junwu and Wen, Sheng and Xiang, Yang},
title = {AI Agents Under Threat: A Survey of Key Security Challenges and Future Pathways},
year = {2025},
issue_date = {July 2025},
publisher = {Association for Computing Machinery},
address = {New York, NY, USA},
volume = {57},
number = {7},
issn = {0360-0300},
url = {https://doi.org/10.1145/3716628},
doi = {10.1145/3716628},
journal = {ACM Comput. Surv.},
month = feb,
articleno = {182},
numpages = {36}
}

@misc{cursor,
  author = {{Anysphere, Inc.}},
  title = {Cursor: The AI Code Editor},
  howpublished = {https://cursor.com},
  year={2026}
}

@online{insta_agent_incident,
    url={https://www.nytimes.com/2026/06/09/technology/instagram-hack-ai-bug.html},
    title={In A.I. Blunder, More Than 34,000 Instagram Accounts Were Attacked},
    author={Mike Isaac and Eli Tan},
    urldate={2026-06-17},
    year={2026}

}

@misc{sierra2026,
  author       = {{Sierra}},
  title        = {Meet your agent},
  year         = {2026},
  howpublished = {\url{https://sierra.ai/product/meet-your-agent}},
  note         = {Accessed: 2026-03-12}
}

@misc{hippocratic2026,
  author       = {{Hippocratic AI}},
  title        = {Hippocratic AI: Home},
  year         = {2026},
  howpublished = {\url{https://hippocraticai.com/}}
}

@misc{openclaw2026,
  author       = {{OpenClaw}},
  title        = {OpenClaw --- Personal AI Assistant},
  year         = {2026},
  howpublished = {\url{https://openclaw.ai/}},
  note         = {Accessed: 2026-03-12}
}

@inproceedings{ouyang2022training,
 author = {Ouyang, Long and Wu, Jeffrey and Jiang, Xu and Almeida, Diogo and Wainwright, Carroll and Mishkin, Pamela and Zhang, Chong and Agarwal, Sandhini and Slama, Katarina and Ray, Alex and Schulman, John and Hilton, Jacob and Kelton, Fraser and Miller, Luke and Simens, Maddie and Askell, Amanda and Welinder, Peter and Christiano, Paul F and Leike, Jan and Lowe, Ryan},
 booktitle = {Advances in Neural Information Processing Systems},
 editor ={S. Koyejo and S. Mohamed and A. Agarwal and D. Belgrave and K. Cho and A. Oh},
 pages = {27730--27744},
 publisher = {Curran Associates, Inc.},
 title = {Training language models to follow instructions with human feedback},
 no_url = {https://proceedings.neurips.cc/paper_files/paper/2022/file/b1efde53be364a73914f58805a001731-Paper-Conference.pdf},
 volume = {35},
 year = {2022}
}

@misc{bai2022hh,
      title={Training a Helpful and Harmless Assistant with Reinforcement Learning from Human Feedback}, 
      author={Yuntao Bai and Andy Jones and Kamal Ndousse and Amanda Askell and Anna Chen and Nova DasSarma and others},
      year={2022},
      eprint={2204.05862},
      archivePrefix={arXiv},
      primaryClass={cs.CL},
      url={https://arxiv.org/abs/2204.05862}, 
}

@inproceedings{perez2022red,
    title = "Red Teaming Language Models with Language Models",
    author = "Perez, Ethan  and
      Huang, Saffron  and
      Song, Francis  and
      Cai, Trevor  and
      Ring, Roman  and
      Aslanides, John  and
      Glaese, Amelia  and
      McAleese, Nat  and
      Irving, Geoffrey",
    editor ="Goldberg, Yoav  and
      Kozareva, Zornitsa  and
      Zhang, Yue",
    booktitle = "Proceedings of the 2022 Conference on Empirical Methods in Natural Language Processing",
    month = dec,
    year = "2022",
    address = "Abu Dhabi, United Arab Emirates",
    publisher = "Association for Computational Linguistics",
    no_url = "https://aclanthology.org/2022.emnlp-main.225/",
    doi = "10.18653/v1/2022.emnlp-main.225",
    pages = "3419--3448",
}

@misc{ganguli2022reduceharms,
      title={Red Teaming Language Models to Reduce Harms: Methods, Scaling Behaviors, and Lessons Learned}, 
      author={Deep Ganguli and Liane Lovitt and Jackson Kernion and Amanda Askell and Yuntao Bai and Saurav Kadavath and others},
      year={2022},
      eprint={2209.07858},
      archivePrefix={arXiv},
      primaryClass={cs.CL},
      url={https://arxiv.org/abs/2209.07858}, 
}

@misc{bai2022constitutional,
      title={Constitutional AI: Harmlessness from AI Feedback}, 
      author={Yuntao Bai and Saurav Kadavath and Sandipan Kundu and Amanda Askell and Jackson Kernion and Andy Jones and others},
      year={2022},
      eprint={2212.08073},
      archivePrefix={arXiv},
      primaryClass={cs.CL},
      url={https://arxiv.org/abs/2212.08073}, 
}

@inproceedings{stiennon2020learning,
 author = {Stiennon, Nisan and Ouyang, Long and Wu, Jeffrey and Ziegler, Daniel and Lowe, Ryan and Voss, Chelsea and Radford, Alec and Amodei, Dario and Christiano, Paul F},
 booktitle = {Advances in Neural Information Processing Systems},
 editor ={H. Larochelle and M. Ranzato and R. Hadsell and M.F. Balcan and H. Lin},
 pages = {3008--3021},
 publisher = {Curran Associates, Inc.},
 title = {Learning to summarize with human feedback},
 no_url = {https://proceedings.neurips.cc/paper_files/paper/2020/file/1f89885d556929e98d3ef9b86448f951-Paper.pdf},
 volume = {33},
 year = {2020}
}

@misc{glaese2022improvingalignmentdialogueagents,
      title={Improving alignment of dialogue agents via targeted human judgements}, 
      author={Amelia Glaese and Nat McAleese and Maja Trębacz and John Aslanides and Vlad Firoiu and Timo Ewalds and Maribeth Rauh and Laura Weidinger and Martin Chadwick and Phoebe Thacker and Lucy Campbell-Gillingham and Jonathan Uesato and Po-Sen Huang and Ramona Comanescu and Fan Yang and Abigail See and Sumanth Dathathri and Rory Greig and Charlie Chen and Doug Fritz and Jaume Sanchez Elias and Richard Green and Soňa Mokrá and Nicholas Fernando and Boxi Wu and Rachel Foley and Susannah Young and Iason Gabriel and William Isaac and John Mellor and Demis Hassabis and Koray Kavukcuoglu and Lisa Anne Hendricks and Geoffrey Irving},
      year={2022},
      eprint={2209.14375},
      archivePrefix={arXiv},
      primaryClass={cs.LG},
      url={https://arxiv.org/abs/2209.14375}, 
}

@inproceedings{lee2023rlaif,
author = {Lee, Harrison and Phatale, Samrat and Mansoor, Hassan and Mesnard, Thomas and Ferret, Johan and Lu, Kellie and Bishop, Colton and Hall, Ethan and Carbune, Victor and Rastogi, Abhinav and Prakash, Sushant},
title = {RLAIF vs. RLHF: scaling reinforcement learning from human feedback with AI feedback},
year = {2024},
publisher = {JMLR.org},
booktitle = {Proceedings of the 41st International Conference on Machine Learning},
articleno = {1071},
numpages = {28},
location = {Vienna, Austria},
series = {ICML'24}
}

@inproceedings{yuan2024selfrewarding,
author = {Yuan, Weizhe and Pang, Richard Yuanzhe and Cho, Kyunghyun and Li, Xian and Sukhbaatar, Sainbayar and Xu, Jing and Weston, Jason},
title = {Self-rewarding language models},
year = {2024},
publisher = {JMLR.org},
booktitle = {Proceedings of the 41st International Conference on Machine Learning},
articleno = {2389},
numpages = {19},
location = {Vienna, Austria},
series = {ICML'24}
}

@inproceedings{rebedea-etal-2023-nemo,
    title = "{N}e{M}o Guardrails: A Toolkit for Controllable and Safe {LLM} Applications with Programmable Rails",
    author = "Rebedea, Traian  and
      Dinu, Razvan  and
      Sreedhar, Makesh Narsimhan  and
      Parisien, Christopher  and
      Cohen, Jonathan",
    editor ="Feng, Yansong  and
      Lefever, Els",
    booktitle = "Proceedings of the 2023 Conference on Empirical Methods in Natural Language Processing: System Demonstrations",
    month = dec,
    year = "2023",
    address = "Singapore",
    publisher = "Association for Computational Linguistics",
    no_url = "https://aclanthology.org/2023.emnlp-demo.40/",
    doi = "10.18653/v1/2023.emnlp-demo.40",
    pages = "431--445"
}

@misc{abaev2026agentguardianlearningaccesscontrol,
      title={AgentGuardian: Learning Access Control Policies to Govern AI Agent Behavior}, 
      author={Nadya Abaev and Denis Klimov and Gerard Levinov and David Mimran and Yuval Elovici and Asaf Shabtai},
      year={2026},
      eprint={2601.10440},
      archivePrefix={arXiv},
      primaryClass={cs.CR},
      url={https://arxiv.org/abs/2601.10440}, 
}

@misc{xiang2025guardagentsafeguardllmagents,
      title={GuardAgent: Safeguard LLM Agents by a Guard Agent via Knowledge-Enabled Reasoning}, 
      author={Zhen Xiang and Linzhi Zheng and Yanjie Li and Junyuan Hong and Qinbin Li and Han Xie and Jiawei Zhang and Zidi Xiong and Chulin Xie and Carl Yang and Dawn Song and Bo Li},
      year={2025},
      eprint={2406.09187},
      archivePrefix={arXiv},
      primaryClass={cs.LG},
      url={https://arxiv.org/abs/2406.09187}, 
}

@inproceedings{luo-etal-2025-agrail,
    title = "{AG}rail: A Lifelong Agent Guardrail with Effective and Adaptive Safety Detection",
    author = "Luo, Weidi  and
      Dai, Shenghong  and
      Liu, Xiaogeng  and
      Banerjee, Suman  and
      Sun, Huan  and
      Chen, Muhao  and
      Xiao, Chaowei",
    editor ="Che, Wanxiang  and
      Nabende, Joyce  and
      Shutova, Ekaterina  and
      Pilehvar, Mohammad Taher",
    booktitle = "Proceedings of the 63rd Annual Meeting of the Association for Computational Linguistics (Volume 1: Long Papers)",
    month = jul,
    year = "2025",
    address = "Vienna, Austria",
    publisher = "Association for Computational Linguistics",
    no_url = "https://aclanthology.org/2025.acl-long.399/",
    doi = "10.18653/v1/2025.acl-long.399",
    pages = "8104--8139",
    ISBN = "979-8-89176-251-0"
}

@misc{chennabasappa2025llamafirewallopensourceguardrail,
      title={LlamaFirewall: An open source guardrail system for building secure AI agents}, 
      author={Sahana Chennabasappa and Cyrus Nikolaidis and Daniel Song and David Molnar and Stephanie Ding and Shengye Wan and Spencer Whitman and Lauren Deason and Nicholas Doucette and Abraham Montilla and Alekhya Gampa and Beto de Paola and Dominik Gabi and James Crnkovich and Jean-Christophe Testud and Kat He and Rashnil Chaturvedi and Wu Zhou and Joshua Saxe},
      year={2025},
      eprint={2505.03574},
      archivePrefix={arXiv},
      primaryClass={cs.CR},
      url={https://arxiv.org/abs/2505.03574}, 
}

@InProceedings{shieldagent,
  title = 	 {{S}hield{A}gent: Shielding Agents via Verifiable Safety Policy Reasoning},
  author =       {Chen, Zhaorun and Kang, Mintong and Li, Bo},
  booktitle = 	 {Proceedings of the 42nd International Conference on Machine Learning},
  pages = 	 {8313--8344},
  year = 	 {2025},
  editor =	 {Singh, Aarti and Fazel, Maryam and Hsu, Daniel and Lacoste-Julien, Simon and Berkenkamp, Felix and Maharaj, Tegan and Wagstaff, Kiri and Zhu, Jerry},
  volume = 	 {267},
  series = 	 {Proceedings of Machine Learning Research},
  month = 	 {13--19 Jul},
  publisher =    {PMLR},
  no_url = 	 {https://proceedings.mlr.press/v267/chen25ae.html}
}

@misc{wang2025agentspeccustomizableruntimeenforcement,
      title={AgentSpec: Customizable Runtime Enforcement for Safe and Reliable LLM Agents}, 
      author={Haoyu Wang and Christopher M. Poskitt and Jun Sun},
      year={2025},
      eprint={2503.18666},
      archivePrefix={arXiv},
      primaryClass={cs.AI},
      url={https://arxiv.org/abs/2503.18666}, 
}

@misc{kamath2025enforcingtemporalconstraintsllm,
      title={Enforcing Temporal Constraints for LLM Agents}, 
      author={Adharsh Kamath and Sishen Zhang and Calvin Xu and Shubham Ugare and Gagandeep Singh and Sasa Misailovic},
      year={2025},
      eprint={2512.23738},
      archivePrefix={arXiv},
      primaryClass={cs.PL},
      url={https://arxiv.org/abs/2512.23738}, 
}

@misc{zhong2025rtbasdefendingllmagents,
      title={RTBAS: Defending LLM Agents Against Prompt Injection and Privacy Leakage}, 
      author={Peter Yong Zhong and Siyuan Chen and Ruiqi Wang and McKenna McCall and Ben L. Titzer and Heather Miller and Phillip B. Gibbons},
      year={2025},
      eprint={2502.08966},
      archivePrefix={arXiv},
      primaryClass={cs.CR},
      url={https://arxiv.org/abs/2502.08966}, 
}

@inproceedings{Halfond2006ACO,
  title={A Classification of SQL Injection Attacks and Countermeasures},
  author={William G. J. Halfond and Jeremy Viegas and Alessandro Orso},
  booktitle={International Symposium on Signals, Systems, and Electronics},
  year={2006},
  url={https://api.semanticscholar.org/CorpusID:5969227}
}

@article{10.1145/360051.360056,
author = {Denning, Dorothy E.},
title = {A lattice model of secure information flow},
year = {1976},
issue_date = {May 1976},
publisher = {Association for Computing Machinery},
address = {New York, NY, USA},
volume = {19},
number = {5},
issn = {0001-0782},
url = {https://doi.org/10.1145/360051.360056},
doi = {10.1145/360051.360056},
journal = {Commun. ACM},
month = may,
pages = {236–243},
numpages = {8}
}

@inproceedings{sandhu2000nist,
  title={The NIST model for role-based access control: towards a unified standard},
  author={Sandhu, Ravi and Ferraiolo, David and Kuhn, Richard},
  booktitle={ACM workshop on Role-based access control},
  volume={10},
  number={344287.344301},
  year={2000}
}

@misc{costa2025securingaiagentsinformationflow,
      title={Securing AI Agents with Information-Flow Control}, 
      author={Manuel Costa and Boris Köpf and Aashish Kolluri and Andrew Paverd and Mark Russinovich and Ahmed Salem and Shruti Tople and Lukas Wutschitz and Santiago Zanella-Béguelin},
      year={2025},
      eprint={2505.23643},
      archivePrefix={arXiv},
      primaryClass={cs.CR},
      url={https://arxiv.org/abs/2505.23643}, 
}

@misc{wu2024systemleveldefenseindirectprompt,
      title={System-Level Defense against Indirect Prompt Injection Attacks: An Information Flow Control Perspective}, 
      author={Fangzhou Wu and Ethan Cecchetti and Chaowei Xiao},
      year={2024},
      eprint={2409.19091},
      archivePrefix={arXiv},
      primaryClass={cs.CR},
      url={https://arxiv.org/abs/2409.19091}, 
}

@misc{debenedetti2025defeatingpromptinjectionsdesign,
      title={Defeating Prompt Injections by Design}, 
      author={Edoardo Debenedetti and Ilia Shumailov and Tianqi Fan and Jamie Hayes and Nicholas Carlini and Daniel Fabian and Christoph Kern and Chongyang Shi and Andreas Terzis and Florian Tramèr},
      year={2025},
      eprint={2503.18813},
      archivePrefix={arXiv},
      primaryClass={cs.CR},
      url={https://arxiv.org/abs/2503.18813}, 
}

@misc{kim2025promptflowintegrityprevent,
      title={Prompt Flow Integrity to Prevent Privilege Escalation in LLM Agents}, 
      author={Juhee Kim and Woohyuk Choi and Byoungyoung Lee},
      year={2025},
      eprint={2503.15547},
      archivePrefix={arXiv},
      primaryClass={cs.CR},
      url={https://arxiv.org/abs/2503.15547}, 
}

@misc{shi2025progentprogrammableprivilegecontrol,
      title={Progent: Programmable Privilege Control for LLM Agents}, 
      author={Tianneng Shi and Jingxuan He and Zhun Wang and Hongwei Li and Linyu Wu and Wenbo Guo and Dawn Song},
      year={2025},
      eprint={2504.11703},
      archivePrefix={arXiv},
      primaryClass={cs.CR},
      url={https://arxiv.org/abs/2504.11703}, 
}

@misc{cui2026marisformallyverifiableprivacy,
      title={Maris: A Formally Verifiable Privacy Policy Enforcement Paradigm for Multi-Agent Collaboration Systems}, 
      author={Jian Cui and Zichuan Li and Luyi Xing and Xiaojing Liao},
      year={2026},
      eprint={2505.04799},
      archivePrefix={arXiv},
      primaryClass={cs.CR},
      url={https://arxiv.org/abs/2505.04799}, 
}

@misc{doshi2026verifiablysafetooluse,
      title={Towards Verifiably Safe Tool Use for LLM Agents}, 
      author={Aarya Doshi and Yining Hong and Congying Xu and Eunsuk Kang and Alexandros Kapravelos and Christian Kästner},
      year={2026},
      eprint={2601.08012},
      archivePrefix={arXiv},
      primaryClass={cs.SE},
      url={https://arxiv.org/abs/2601.08012}, 
}

@misc{palumbo2026policycompilersecureagentic,
      title={Policy Compiler for Secure Agentic Systems}, 
      author={Nils Palumbo and Sarthak Choudhary and Jihye Choi and Prasad Chalasani and Somesh Jha},
      year={2026},
      eprint={2602.16708},
      archivePrefix={arXiv},
      primaryClass={cs.CR},
      url={https://arxiv.org/abs/2602.16708}, 
}

@inproceedings{felizardo2016using,
  title={Using forward snowballing to update systematic reviews in software engineering},
  author={Felizardo, Katia Romero and Mendes, Emilia and Kalinowski, Marcos and Souza, {\'E}rica Ferreira and Vijaykumar, Nandamudi L},
  booktitle={Proceedings of the 10th ACM/IEEE International Symposium on Empirical Software Engineering and Measurement},
  pages={1--6},
  year={2016}
}

@misc{chatgpt,
  author       = {{OpenAI}},
  title        = {ChatGPT},
  year         = {2026},
  howpublished = {\url{https://chatgpt.com/}},
  note         = {Large language model accessed March 20, 2026}
}

@misc{yao2024tau,
      title={$\tau$-bench: A Benchmark for Tool-Agent-User Interaction in Real-World Domains}, 
      author={Shunyu Yao and Noah Shinn and Pedram Razavi and Karthik Narasimhan},
      year={2024},
      eprint={2406.12045},
      archivePrefix={arXiv},
      primaryClass={cs.AI},
      url={https://arxiv.org/abs/2406.12045}, 
}

@inproceedings{
levy2026stwebagentbench,
title={{ST}-WebAgentBench: A Benchmark for Evaluating Safety and Trustworthiness in Web Agents},
author={Ido Levy and Ben wiesel and Sami Marreed and Alon Oved and Avi Yaeli and Segev Shlomov},
booktitle={The Fourteenth International Conference on Learning Representations},
year={2026},
url={https://openreview.net/forum?id=MuCDzH0ctf}
}

@misc{lei2026offtopicevallargelanguagemodels,
      title={OffTopicEval: When Large Language Models Enter the Wrong Chat, Almost Always!}, 
      author={Jingdi Lei and Varun Gumma and Rishabh Bhardwaj and Seok Min Lim and Chuan Li and Amir Zadeh and Soujanya Poria},
      year={2026},
      eprint={2509.26495},
      archivePrefix={arXiv},
      primaryClass={cs.AI},
      url={https://arxiv.org/abs/2509.26495}, 
}

@misc{chen2026lpsbenchbenchmarkingsafetyawareness,
      title={LPS-Bench: Benchmarking Safety Awareness of Computer-Use Agents in Long-Horizon Planning under Benign and Adversarial Scenarios}, 
      author={Tianyu Chen and Chujia Hu and Ge Gao and Dongrui Liu and Xia Hu and Wenjie Wang},
      year={2026},
      eprint={2602.03255},
      archivePrefix={arXiv},
      primaryClass={cs.AI},
      url={https://arxiv.org/abs/2602.03255}, 
}

@article{Lee_Hahm_Choi_Knox_Lee_2026, title={MobileSafetyBench: Evaluating Safety of Autonomous Agents in Mobile Device Control}, volume={40}, url={https://ojs.aaai.org/index.php/AAAI/article/view/41090}, DOI={10.1609/aaai.v40i44.41090}, abstractNote={Autonomous agents powered by large language models (LLMs) show promising potential in assistive tasks across various domains, including mobile device control. As these agents interact directly with personal information and device settings, ensuring their safe and reliable behavior is crucial to prevent undesirable outcomes. However, no benchmark exists for standardized evaluation of the safety of mobile device-control agents. In this work, we introduce MobileSafetyBench, a benchmark designed to evaluate the safety of device-control agents within a realistic mobile environment based on Android emulators. We develop a diverse set of tasks involving interactions with various mobile applications, including messaging and banking applications, challenging agents with managing risks encompassing the misuse and negative side effects. These tasks include tests to evaluate the safety of agents in daily scenarios as well as their robustness against indirect prompt injection attacks. Our experiments demonstrate that baseline agents, based on state-of-the-art LLMs, often fail to effectively prevent harm while performing the tasks. To mitigate these safety concerns, we propose a prompting method that encourages agents to prioritize safety considerations. While this method shows promise in promoting safer behaviors, there is still considerable room for improvement to fully earn user trust. This highlights the urgent need for continued research to develop more robust safety mechanisms in mobile environments.}, number={44}, journal={Proceedings of the AAAI Conference on Artificial Intelligence}, author={Lee, Juyong and Hahm, Dongyoon and Choi, June Suk and Knox, W. Bradley and Lee, Kimin}, year={2026}, month={Mar.}, pages={37565-37573} }

@misc{zhou2026safeproevaluatingsafetyprofessionallevel,
      title={SafePro: Evaluating the Safety of Professional-Level AI Agents}, 
      author={Kaiwen Zhou and Shreedhar Jangam and Ashwin Nagarajan and Tejas Polu and Suhas Oruganti and Chengzhi Liu and Ching-Chen Kuo and Yuting Zheng and Sravana Narayanaraju and Xin Eric Wang},
      year={2026},
      eprint={2601.06663},
      archivePrefix={arXiv},
      primaryClass={cs.AI},
      url={https://arxiv.org/abs/2601.06663}, 
}

@misc{wang2026mpcibenchbenchmarkmultimodalpairwise,
      title={MPCI-Bench: A Benchmark for Multimodal Pairwise Contextual Integrity Evaluation of Language Model Agents}, 
      author={Shouju Wang and Haopeng Zhang},
      year={2026},
      eprint={2601.08235},
      archivePrefix={arXiv},
      primaryClass={cs.AI},
      url={https://arxiv.org/abs/2601.08235}, 
}

@inproceedings{
andriushchenko2025agentharm,
title={AgentHarm: A Benchmark for Measuring Harmfulness of {LLM} Agents},
author={Maksym Andriushchenko and Alexandra Souly and Mateusz Dziemian and Derek Duenas and Maxwell Lin and Justin Wang and Dan Hendrycks and Andy Zou and J Zico Kolter and Matt Fredrikson and Yarin Gal and Xander Davies},
booktitle={The Thirteenth International Conference on Learning Representations},
year={2025},
url={https://openreview.net/forum?id=AC5n7xHuR1}
}

@misc{kirmayr2026carbenchevaluatingconsistencylimitawareness,
      title={CAR-bench: Evaluating the Consistency and Limit-Awareness of LLM Agents under Real-World Uncertainty}, 
      author={Johannes Kirmayr and Lukas Stappen and Elisabeth André},
      year={2026},
      eprint={2601.22027},
      archivePrefix={arXiv},
      primaryClass={cs.AI},
      url={https://arxiv.org/abs/2601.22027}, 
}

@INPROCEEDINGS{11030006,
  author={Hong, Yining and Timperley, Christopher S. and Kästner, Christian},
  booktitle={2025 IEEE/ACM 4th International Conference on AI Engineering – Software Engineering for AI (CAIN)}, 
  title={From Hazard Identification to Controller Design: Proactive and LLM-Supported Safety Engineering for ML-Powered Systems}, 
  year={2025},
  volume={},
  number={},
  pages={113-118},
  doi={10.1109/CAIN66642.2025.00021}}

@misc{cuadron2025sabersmallactionsbig,
      title={SABER: Small Actions, Big Errors - Safeguarding Mutating Steps in LLM Agents}, 
      author={Alejandro Cuadron and Pengfei Yu and Yang Liu and Arpit Gupta},
      year={2025},
      eprint={2512.07850},
      archivePrefix={arXiv},
      primaryClass={cs.LG},
      url={https://arxiv.org/abs/2512.07850}, 
}

@online{anthropic_claude_code,
  author       = {{Anthropic}},
  title        = {Claude Code Overview},
  year         = {2026},
  url          = {https://docs.anthropic.com/en/docs/claude-code/overview},
  note         = {Accessed: 2026-03-25}
}

@inproceedings {299563,
author = {Yupei Liu and Yuqi Jia and Runpeng Geng and Jinyuan Jia and Neil Zhenqiang Gong},
title = {Formalizing and Benchmarking Prompt Injection Attacks and Defenses},
booktitle = {33rd USENIX Security Symposium (USENIX Security 24)},
year = {2024},
isbn = {978-1-939133-44-1},
address = {Philadelphia, PA},
pages = {1831--1847},
url = {https://www.usenix.org/conference/usenixsecurity24/presentation/liu-yupei},
publisher = {USENIX Association},
month = aug
}

@techreport{kitchenham2007guidelines,
  title={Guidelines for performing systematic literature reviews in software engineering},
  author={Kitchenham, Barbara and Charters, Stuart},
  year={2007},
  publisher={Keele, UK},
  institution = {Keele University and University of Durham}
}

@inproceedings{scholak-etal-2021-picard,
    title = "{PICARD}: Parsing Incrementally for Constrained Auto-Regressive Decoding from Language Models",
    author = "Scholak, Torsten  and
      Schucher, Nathan  and
      Bahdanau, Dzmitry",
    editor = "Moens, Marie-Francine  and
      Huang, Xuanjing  and
      Specia, Lucia  and
      Yih, Scott Wen-tau",
    booktitle = "Proceedings of the 2021 Conference on Empirical Methods in Natural Language Processing",
    month = nov,
    year = "2021",
    address = "Online and Punta Cana, Dominican Republic",
    publisher = "Association for Computational Linguistics",
    no_url = "https://aclanthology.org/2021.emnlp-main.779/",
    doi = "10.18653/v1/2021.emnlp-main.779",
    pages = "9895--9901"
}

@book{viega2001building,
  title={Building secure software: how to avoid security problems the right way},
  author={Viega, John and McGraw, Gary R},
  year={2001},
  publisher={Pearson Education}
}

@misc{yang2025promptsdontsayunderstanding,
      title={What Prompts Don't Say: Understanding and Managing Underspecification in LLM Prompts}, 
      author={Chenyang Yang and Yike Shi and Qianou Ma and Michael Xieyang Liu and Christian Kästner and Tongshuang Wu},
      year={2025},
      eprint={2505.13360},
      archivePrefix={arXiv},
      primaryClass={cs.CL},
      url={https://arxiv.org/abs/2505.13360}, 
}

@misc{1335256,
  author = {G Stoneburner and Alice Goguen and Alexis Feringa},
  title = {Risk Management Guide for Information Technology Systems},
  year = {2002},
  month = {2002-07-01 00:07:00},
  publisher = {Special Publication (NIST SP), National Institute of Standards and Technology, Gaithersburg, MD},
  url = {https://tsapps.nist.gov/publication/get_pdf.cfm?pub_id=151254},
  doi = {https://doi.org/10.6028/nist.sp.800-30},
  language = {en},
}

@misc{dehghani2021benchmarklottery,
      title={The Benchmark Lottery}, 
      author={Mostafa Dehghani and Yi Tay and Alexey A. Gritsenko and Zhe Zhao and Neil Houlsby and Fernando Diaz and Donald Metzler and Oriol Vinyals},
      year={2021},
      eprint={2107.07002},
      archivePrefix={arXiv},
      primaryClass={cs.LG},
      url={https://arxiv.org/abs/2107.07002}, 
}

@inproceedings{
raji2021ai,
title={{AI} and the Everything in the Whole Wide World Benchmark},
author={Inioluwa Deborah Raji and Emily Denton and Emily M. Bender and Alex Hanna and Amandalynne Paullada},
booktitle={Thirty-fifth Conference on Neural Information Processing Systems Datasets and Benchmarks Track (Round 2)},
year={2021},
no_url={https://openreview.net/forum?id=j6NxpQbREA1}
}

@online{Cecco2024AirCanadaChatbot,
  author  = {Cecco, Leyland},
  title   = {Air Canada ordered to pay customer who was misled by airline's chatbot},
  year    = {2024},
  date    = {2024-02-16},
  url     = {https://www.theguardian.com/world/2024/feb/16/air-canada-chatbot-lawsuit},
  urldate = {2026-06-26},
  organization = {The Guardian},
}

@misc{tan2025equilibraterlhfbalancinghelpfulnesssafety,
      title={Equilibrate RLHF: Towards Balancing Helpfulness-Safety Trade-off in Large Language Models}, 
      author={Yingshui Tan and Yilei Jiang and Yanshi Li and Jiaheng Liu and Xingyuan Bu and Wenbo Su and Xiangyu Yue and Xiaoyong Zhu and Bo Zheng},
      year={2025},
      eprint={2502.11555},
      archivePrefix={arXiv},
      primaryClass={cs.AI},
      url={https://arxiv.org/abs/2502.11555}, 
}

@article{dong2025safeguarding,
  title={Safeguarding large language models: A survey},
  author={Dong, Yi and Mu, Ronghui and Zhang, Yanghao and Sun, Siqi and Zhang, Tianle and Wu, Changshun and Jin, Gaojie and Qi, Yi and Hu, Jinwei and Meng, Jie and others},
  journal={Artificial intelligence review},
  volume={58},
  number={12},
  pages={382},
  year={2025},
  publisher={Springer}
}

@standard{ISOIECIEEE29148,
  title        = {{ISO/IEC/IEEE International Standard -- Systems and software engineering -- Life cycle processes -- Requirements engineering}},
  organization = {{ISO/IEC/IEEE}},
  number       = {{ISO/IEC/IEEE 29148:2018}},
  year         = {2018},
  month        = nov,
  publisher    = {{IEEE}},
  doi          = {10.1109/IEEESTD.2018.8559686}
}

@techreport{Berry2003Ambiguity,
  author      = {Berry, Daniel M. and Kamsties, Erik and Krieger, Michael M.},
  title       = {{From Contract Drafting to Software Specification: Linguistic Sources of Ambiguity}},
  institution = {{University of Waterloo}},
  year        = {2003},
  month       = nov,
  url         = {https://cs.uwaterloo.ca/~dberry/handbook/ambiguityHandbook.pdf}
}

@article{FEMMER2017190,
title = {Rapid quality assurance with Requirements Smells},
journal = {Journal of Systems and Software},
volume = {123},
pages = {190-213},
year = {2017},
issn = {0164-1212},
doi = {https://doi.org/10.1016/j.jss.2016.02.047},
no_url = {https://www.sciencedirect.com/science/article/pii/S0164121216000789},
author = {Henning Femmer and Daniel {Méndez Fernández} and Stefan Wagner and Sebastian Eder}
}

@ARTICLE{8254303,
  author={Femmer, Henning and Vogelsang, Andreas},
  journal={IEEE Software}, 
  title={Requirements Quality Is Quality in Use}, 
  year={2019},
  volume={36},
  number={3},
  pages={83-91},
  doi={10.1109/MS.2018.110161823}}

@book{Engineering_a_safer_world_Nancy2012,
    author = {Leveson, Nancy G.},
    title = "{Engineering a Safer World: Systems Thinking Applied to Safety}",
    publisher = {The MIT Press},
    year = {2012},
    month = {01},
    isbn = {9780262298247},
    doi = {10.7551/mitpress/8179.001.0001}
}

@article{sabelfeld2003language,
  title={Language-based information-flow security},
  author={Sabelfeld, Andrei and Myers, Andrew C},
  journal={IEEE Journal on selected areas in communications},
  volume={21},
  number={1},
  pages={5--19},
  year={2003},
  publisher={IEEE}
}
\bibliographystyle{IEEEtran}

\end{document}